\documentclass[9pt,twocolumn,twoside]{pnas-new}

\usepackage{amsmath}
\usepackage{graphicx}
\usepackage{dcolumn}
\usepackage{bm}
\usepackage{gensymb}
\usepackage{svg}
\usepackage{afterpage}
\usepackage{physics}
\usepackage{tabularx}
\usepackage{caption}
\usepackage{mathtools}

\usepackage{soul}
\usepackage{currfile} 

\usepackage{xr}

\makeatletter
\newcommand*{\addFileDependency}[1]{ 
  \typeout{(#1)}
  \@addtofilelist{#1}
  \IfFileExists{#1}{}{\typeout{No file #1.}}
}
\makeatother

\newcommand*{\myexternaldocument}[1]{%
    \externaldocument{#1}%
    \addFileDependency{#1.aux}%
}

\myexternaldocument{SI}

\templatetype{pnasresearcharticle} 

\begin{document}

\title{Transport and Energetics of Bacterial Rectification}

\author[a,b,c,1]{Satyam Anand}
\author[c]{Xiaolei Ma}
\author[c]{Shuo Guo}
\author[a,b,d,1,*]{Stefano Martiniani}
\author[c,1,*]{Xiang Cheng}

\affil[a]{Courant Institute of Mathematical Sciences, New York University, New York, NY 10003, USA}
\affil[b]{Center for Soft Matter Research, Department of Physics, New York University, New York, NY 10003, USA}
\affil[c]{Department of Chemical Engineering and Materials Science, University of Minnesota, Minneapolis, MN 55455, USA}
\affil[d]{Simons Center for Computational Physical Chemistry, Department of Chemistry, New York University, New York, NY 10003, USA}

\leadauthor{Anand}

\significancestatement{Popularized by Feynman in his renowned lectures, the Brownian ratchet captures the paradox of a rectification process, where directed motion arises spontaneously from microscopic random fluctuations in out-of-equilibrium settings. Rectification occurs ubiquitously in living systems and is essential to many important biological functions. Despite extensive research, however, the transport and energetics of rectification of living systems remain poorly understood. Here, we investigate the rectification of randomly swimming bacteria as they navigate through fixed asymmetric obstacles, a paradigmatic model of rectification of living systems. The synergy of our experimental, numerical, and theoretical results provides a quantitative description of bacterial rectification based on single-bacterium dynamics and illustrates the generic principles governing the energetics of this non-equilibrium process.} 

\authorcontributions{S.A., S.M. and X.C. conceived the project and designed the research. S.A. conducted PDMS microfluidic channel experiments, performed simulations, and developed the theory. X.M. and S.G. conducted the optical tweezers experiments. S.A., S.M., and X.C. analyzed data and wrote the manuscript. S.M. and X.C. supervised the research.}
\authordeclaration{The authors declare no competing interest.}
\correspondingauthor{\textsuperscript{*}Equal Contribution. \\
\textsuperscript{1}To whom correspondence should be addressed. E-mail: sa7483@nyu.edu, sm7683@nyu.edu, or xcheng@umn.edu}


\begin{abstract}
Randomly moving active particles can be herded into directed motion by asymmetric geometric structures. Although such a rectification process has been extensively studied due to its fundamental, biological, and technological relevance, a comprehensive understanding of active matter rectification based on single particle dynamics remains elusive. Here, by combining experiments, simulations, and theory, we study the directed transport and energetics of swimming bacteria navigating through funnel-shaped obstacles---a paradigmatic model of rectification of living active matter. We develop a microscopic parameter-free model for bacterial rectification, which quantitatively explains experimental and numerical observations and predicts the optimal geometry for the maximum rectification efficiency. Furthermore, we quantify the degree of time irreversibility and measure the extractable work associated with bacterial rectification. Our study provides quantitative solutions to long-standing questions on bacterial rectification and establishes a generic relationship between time irreversibility, particle fluxes, and extractable work, shedding light on the energetics of non-equilibrium rectification processes in living systems.    
\end{abstract}

\dates{This manuscript was compiled on \today}
\doi{\url{www.pnas.org/cgi/doi/10.1073/pnas.XXXXXXXXXX}}

\maketitle
\thispagestyle{firststyle}
\ifthenelse{\boolean{shortarticle}}{\ifthenelse{\boolean{singlecolumn}}{\abscontentformatted}{\abscontent}}{}

\firstpage[3]{3}

\dropcap{R}ectification of microscopic fluctuations to create directed transport has been a topic of interest for well over a century, with one of the first examples being the Brownian or Feynman-Smoluchowski ratchet \cite{feynman1971feynman, smoluchowski1912experimental}. The process is particularly important for living systems, where various types of molecular motors need to operate persistently in a unidirectional fashion against thermal noise \cite{yuan2021fundamentals}. To rectify a thermal system, both the time-reversal and spatial inversion symmetries of the system must be broken \cite{reimann2002brownian}, which necessarily requires the input of energy from either internal or external sources. For a typical biological process such as the unidirectional movement of kinesins along microtubules, the breaking of the two symmetries is often complicated by numerous physical, chemical, and biological factors, making it challenging to dissect the detailed dynamics and energetics of rectification in living systems \cite{ariga2018nonequilibrium, ariga2020experimental}.

Composed of independent self-propelled units, active matter provides a paradigm for the study of symmetry-breaking mechanisms and non-equilibrium dynamics in rectified living systems under well controlled conditions. The self-propulsion of active particles intrinsically breaks time-reversal symmetry, whereas the breaking of spatial inversion symmetry can be achieved independently by using asymmetric geometric structures \cite{angelani2009self, di2010bacterial, kaiser2014transport, li2013asymmetric, pietzonka2019autonomous, sokolov2010swimming, angelani2010geometrically}. Notably, in seminal experiments on swimming bacteria, Austin, Chaikin and co-workers found that microfluidic devices with an array of fixed funnel-shaped obstacles spontaneously induce a concentration difference of active particles between the two sides of the array from an initially uniform active bath \cite{galajda2007wall}. The simplicity of the bacterial rectification process, together with the great potential in biotechnological applications such as cell sorting and trapping \cite{galajda2007wall, drocco2012bidirectional, martinez2020trapping, guidobaldi2014geometrical}, micro-patterning \cite{galajda2007wall}, spontaneous pumping \cite{galajda2007wall} and cargo transport \cite{koumakis2013targeted}, has stimulated extensive studies on the rectification of active matter in recent years \cite{reichhardt2017ratchet}, both numerically \cite{wan2008rectification, reichhardt2011active, tailleur2009sedimentation, ro2022model} and in experiments \cite{galajda2007wall, galajda2008funnel, lambert2010collective, ro2022model, kantsler2013ciliary, sparacino2020solitary, nam2013c, guidobaldi2014geometrical, galajda2008funnel}. Although it is generally acknowledged that bacterial rectification arises from bacterium-obstacle interactions \cite{galajda2007wall}, a quantitative understanding of the directed transport based on the microscopic dynamics of bacteria has not yet been achieved. Despite intensive research over the last two decades, important questions such as the identification of the geometry with the maximum rectification efficiency remain unresolved. Answers to these questions may shed light on the evolution of carnivorous plants that rectify the motion of microbes through root hairs structurally similar to funnel-shaped obstacles \cite{martin2023carnivorous}.

Furthermore, most of the existing studies of bacterial rectification focus on directed transport \cite{wan2008rectification, reichhardt2011active, tailleur2009sedimentation, ro2022model, galajda2007wall, galajda2008funnel, lambert2010collective, ro2022model}. The energetics underlying this non-equilibrium process have not been explored. Several questions regarding the energetics of bacterial rectification are particularly interesting. First, while it is clear that bacteria consume chemical energy to move and break time-reversal symmetry, how much additional energy, if any, is required to rectify bacterial motion with \textit{fixed} passive structures (viz., obstacles)? Second, how much useful work can be harnessed from the rectified bacterial motion? Can extractable work be measured with minimal perturbations to particle transport and the degree of time-reversal symmetry breaking of the system? Finally, what are the quantitative relations between the basic thermodynamic quantities such as particle flux, time irreversibility, and extractable work? The answers to these questions would establish the foundation for comprehending the non-equilibrium thermodynamics of the rectification process and illuminate the overarching principles governing the energetics of rectified active systems.

Here, by combining experiments, simulations and theory, we provide a quantitative microscopic understanding of the transport and energetics of bacterial rectification---a paradigmatic model of the rectification of living active matter \cite{reichhardt2017ratchet, cates2012diffusive}. In particular, we identify three essential factors controlling bacterial rectification: (i) a universal distribution of the orientation of bacteria that swim across a linear segment such as the opening of obstacles; (ii) the realignment of bacteria along solid boundaries; (iii) bacterial wobbling. A model incorporating these factors quantitatively describes bacterial transport spanning a wide range of obstacle shapes and predicts the optimal geometry that maximizes the rectification efficiency. Furthermore, we characterize the time irreversibility of bacterial rectification in terms of its local entropy production rate (EPR) and analyze the energetics of the process. Finally, we devise a protocol to measure extractable work through a weak coupling to bacterial flux, which allows us to elucidate the intrinsic relation between three fundamental thermodynamics quantities---particle fluxes, EPR, and extractable work, the first of this kind in a single non-equilibrium process. Taken together, our study provides not only important insights into rectification processes in living systems but also a useful guideline for controlling active particles in biotechnological applications \cite{galajda2007wall, drocco2012bidirectional, martinez2020trapping, guidobaldi2014geometrical, koumakis2013targeted}.

In the following, we first show results on the directed transport in bacterial rectification and then analyze the energetics of the process. Finally, we discuss implications of our results and open questions.

\begin{figure}[t!]
    \centering
    \includegraphics[width=\linewidth]{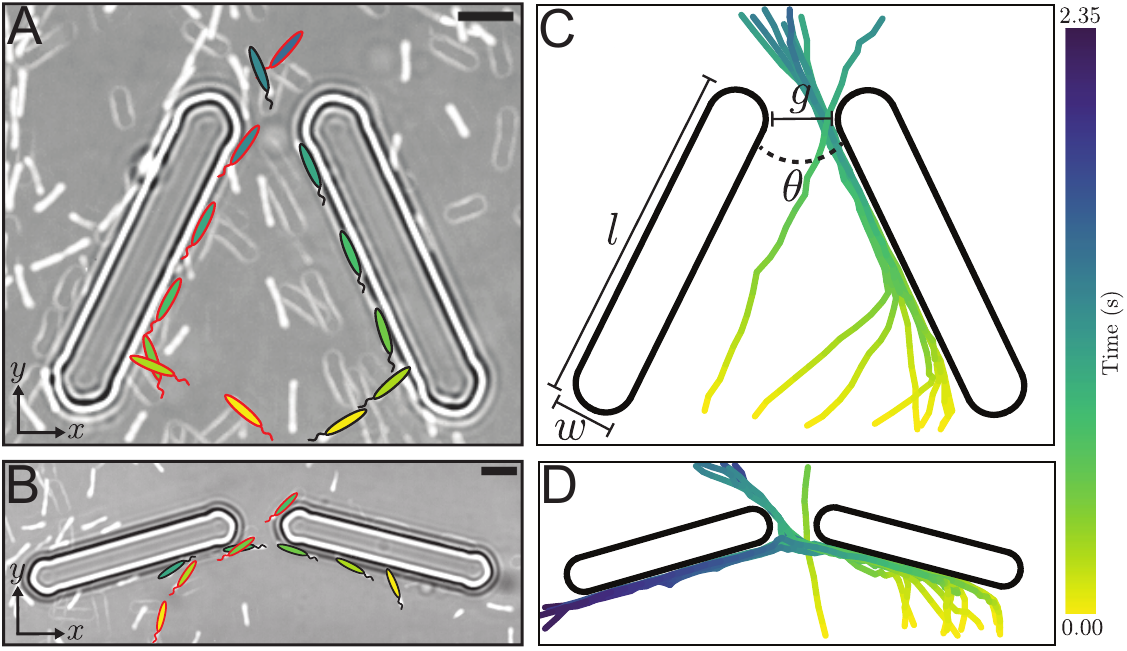}
    \caption{Bacterial rectification induced by bacterium-wall interactions. Superimposed time-series showing rectification of two exemplar \textit{E. coli} by funnels of angle $\theta=50\degree$ (\textit{A}) and $150\degree$ (\textit{B}) (Movies $1$ and $2$). Stylized ellipsoids placed on top of actual cell bodies are color-coded to show time evolution, and the black and red outlines distinguish the two different bacteria. Representative bacterial trajectories for ${\theta=50\degree}$ (\textit{C}) and $150\degree$ (\textit{D}). For clarity, only bacteria rectified by the right funnel wall are shown. The geometrical parameters of the funnels are indicated in (\textit{C}). Scale bars: 5 $\mu$m.}
    \label{fig:figure1}
\end{figure} 

\begin{figure*}[t!]
    \centering
    \includegraphics[width=\linewidth]{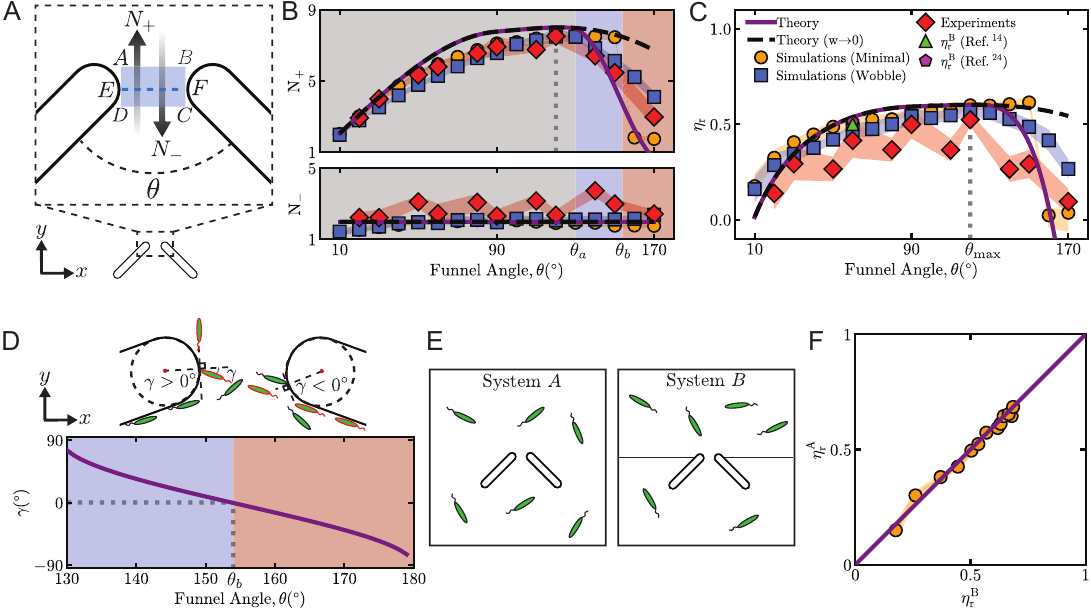}
    \caption{Directed transport of bacterial rectification. (\textit{A}) Schematic showing bacterial fluxes through a funnel tip of angle $\theta$. $EF$ is the center line of the region of interest $ABCD$. ${N}_{+}$ and ${N}_{-}$ indicate normalized number of particles crossing $EF$ per unit length and time in the $+y$ and $-y$ directions, respectively. (\textit{B}) ${N}_{+}$ and ${N}_{-}$ as a function of $\theta$. Different regimes are shaded differently with $\theta_a  = 130 \degree$ and $\theta_b = 153 \degree$ (see the main text). The vertical dotted line marks $\theta_{\max} = 120\degree$. The meaning of symbols and lines are given in the legend of (\textit{C}). (\textit{C}) Rectification efficiency (viz., normalized flux) $\eta_r = (N_+ - N_-)/(N_+ + N_-)$ as a function of $\theta$. Data from previous experiments \cite{galajda2007wall, galajda2008funnel} are extracted for comparison based on Eq.~\ref{eq:c_a_j_b_eta}. Standard deviations of measurements are indicated by the shaded regions around data. (\textit{D}) (Top) Schematics showing the trajectories of bacteria near the funnel tip when $\theta > \theta_a$. The bacterium coming from the left displays wobbling before hitting the wall tip. $\gamma$ is the angle between the bacterial orientation and the local normal of the wall. $\gamma >0$ when a bacterium is below the normal (left). (Bottom) $\gamma$ as a function of $\theta$. $\gamma = 0\degree$ when $\theta = \theta_b$. (\textit{E}) Schematics of our system (System $A$) with non-zero flux $J^A$ at steady state and of System $B$ from previous studies with a finite concentration difference $\Delta C^B$ between the $+y$ and $-y$ regions at steady state. (\textit{F}) The rectification efficiency of System $A$, $\eta^A_r$, versus the rectification efficiency of System $B$ (viz., normalized concentration difference), $\eta^B_r = \Delta C/(2C_0)$. The solid purple line is the prediction of Eq.~\ref{eq:c_a_j_b_eta}.
}
    \label{fig:figure2}
\end{figure*}

\section*{Transport}

\subsection*{Experiments and Simulations} We inject a dilute suspension of \textit{Escherichia coli} (\textit{E. coli}) in a quasi-two-dimensional (2D) polydimethylsiloxane (PDMS) microfluidic chamber containing an isolated funnel-shaped obstacle with angle ($\theta$), length ($l$), width ($w$) and gap ($g$) (Fig.~\ref{fig:figure1}, see also \textit{SI Appendix}, Fig.~\ref{fig:figure1_SI}\textit{A} for the idealized representation adopted in simulations). Unless stated otherwise, $l/g=6$ and $w/g=1$ in our study. The trajectories of bacteria and the interactions between bacteria and funnel walls are imaged via optical microscopy (Fig.~\ref{fig:figure1}, Materials and Methods).

Away from funnel walls, freely swimming \textit{E. coli} exhibit the classic run-and-tumble motion \cite{berg2004coli}. The average run length of bacteria in our quasi-2D chamber is significantly longer than the length of the funnel walls. Notably, the trajectory of a bacterium in the run phase is a 3D helix, which manifests as wobbling of the bacterial body under the 2D projection of optical microscopy \cite{hyon2012wiggling, kamdar2022colloidal} (Movies 1 and 2). Independent of the angle of incidence, a bacterium always re-aligns and moves parallel to the funnel walls after hitting the walls (Figs.~\ref{fig:figure1}\textit{A} and \textit{B}) \cite{galajda2007wall}. In our minimal simulations, we model bacteria as non-interacting point particles performing run-and-tumble motions without wobbling, and incorporate the surface alignment trait of \textit{E. coli} via event-driven dynamics (Materials and Methods).

Our locally defined system of interest is a small region \textit{ABCD} at the funnel tip (Fig.~\ref{fig:figure2}\textit{A}). The number of bacteria crossing the center-line \textit{EF} of \textit{ABCD} per unit time and length in the $+y$ $(-y)$ direction, normalized by the same quantity from a region far from the funnel gap, is denoted by ${N}_{+}$ (${N}_{-}$). Figure~\ref{fig:figure2}\textit{B} shows ${N}_{+}$ and ${N}_{-}$ as a function of $\theta$. Since $w = g$ is small compared with $l$, reverse rectification is not pronounced, leading to a nearly constant ${N}_{-} \approx 1.8 $. In contrast, ${N}_{+}$ shows a clear non-monotonic trend with increasing $\theta$. Minimal simulations show good agreement with experiments for $\theta < \theta_a \approx 130 \degree$ but deviate significantly from experiments when $\theta > \theta_a$ (Fig.~\ref{fig:figure2}\textit{B}). Accordingly, we discuss our results in these two regimes separately.

(\textit{i}) $\theta < \theta_a$. As more bacteria can enter a funnel of wider opening, ${N}_{+}$ increases with $\theta < \theta_\mathrm{max}$ (Fig.~\ref{fig:figure2}\textit{B}). Less intuitively, the number of bacteria exiting the funnel in the $-y$ direction (viz., that rebound) after hitting the interior funnel walls also increases with $\theta$. The latter effect gradually dominates the rectification process with increasing $\theta$, giving rise to the peak of $N_{+}$ at $\theta = \theta_{\max} \approx 120\degree$ in both our experiments and minimal simulations. We illustrate the competition of the two effects quantitatively in a microscopic model below. 

The agreement between our minimal simulations and experiments for $\theta < \theta_a$ has two important implications. First, even though bacteria-wall interactions are governed by complex steric and near-field hydrodynamic forces \cite{berke2008hydrodynamic, li2009accumulation, bianchi2017holographic, drescher2011fluid}, from the perspective of rectification, such complex interactions can be effectively condensed into one simple rule: \textit{E. coli} reorient themselves parallel to the wall after a collision. Such a simple rule has been ignored in most numerical studies of bacterial rectification \cite{wan2008rectification, ro2022model}, with notable exceptions \cite{reichhardt2011active, tailleur2009sedimentation}. Second, the rectification efficiency can be defined as the net flux per particle, $\eta_r \equiv ({N}_{+}-{N}_{-})/({N}_{+}+{N}_{-})$. The non-monotonic trend in ${N}_{+}$, in combination with the nearly constant ${N}_{-}$, results in a maximum $\eta_r$ at $\theta = \theta_{\max}$ (Fig.~\ref{fig:figure2}\textit{C}).

(\textit{ii}) $\theta > \theta_a$. At large enough angles, following the realignment induced by the collision with one funnel wall, bacteria would also interact with the tip of the opposite wall before leaving the funnel (Fig.~\ref{fig:figure2}\textit{D}, \textit{SI Appendix},  Sec. II.A.2, Fig.~\ref{fig:figure1_SI}). This event profoundly modifies the outcome of rectification. The onset of the regime is marked by $\theta_a = 129.22 \degree$, where the trajectory of bacteria becomes tangential to the tip of the opposite wall (\textit{SI Appendix}, Fig.~\ref{fig:figure1_SI}\textit{C}). The angle between the orientation of a bacterium and the local normal of the funnel wall, $\gamma$, dictates whether a bacterium goes in the $+y$ or $-y$ direction after collision with the opposite wall (Fig.~\ref{fig:figure2}\textit{D}, \textit{SI Appendix}, Figs.~\ref{fig:figure1_SI}\textit{C-F}). For $\gamma > 0 \degree$, a bacterium hits the wall below the normal, moves in the $+y$ direction, and is therefore rectified (left schematic in Fig.~\ref{fig:figure2}\textit{D}), whereas for $\gamma < 0 \degree$ it turns back in the $-y$ direction without rectification (right schematic in Fig.~\ref{fig:figure2}\textit{D}). At $\theta = \theta_b = 153.74 \degree$, the trajectory of a bacterium aligns with the local normal of the wall, corresponding to $\gamma = 0 \degree$ (Fig.~\ref{fig:figure2}\textit{D}, \textit{SI Appendix}, Fig.~\ref{fig:figure1_SI}\textit{E}). Thus, $\theta_b$ is a separation angle setting the two different fates of the bacterium apart (Fig.~\ref{fig:figure2}\textit{D}), which explains the sharp change of $N_+$ at $\theta_b$ in minimal simulations (Fig.~\ref{fig:figure2}\textit{B}).

Compared to minimal simulations, the change of $N_+$ in experiments is more gradual (Fig.~\ref{fig:figure2}\textit{B}). This discrepancy arises due to the wobbling of bacteria. Wobbling perturbs the orientation of a bacterium with respect to the local normal at the moment of collision, which can change the sign of $\gamma$ and alter the fortune of the bacterium. The effect is particularly strong for $\theta$ close to $\theta_b$, where even a small amount of wobbling is sufficient to change the sign of $\gamma$ (the right schematic in Fig.~\ref{fig:figure2}\textit{D}, \textit{SI Appendix}, Fig.~\ref{fig:figure1_SI}\textit{E}). The effect of wobbling becomes weaker as $\theta$ deviates from $\theta_b$ on either side (Fig.~\ref{fig:figure2}\textit{B}).

To demonstrate the effect of wobbling on bacterial rectification, we incorporate bacterial wobbling in our minimal simulations by introducing an internal orientation vector for our particles which, when misaligned with the velocity vector, mimics bacterial wobbling (Materials and Methods). Simulations with wobbling bacteria quantitatively match our experiments (Figs.~\ref{fig:figure2}\textit{B} and \textit{C}). Thus, our experiments and simulations show that both the bacterium-wall interaction and bacterial wobbling are important in controlling the directed transport of bacterial rectification in a funnel geometry.

\subsection*{Microscopic Model} Guided by the experimental and numerical findings, we develop an analytical model of bacterial rectification based on microscopic bacterial dynamics. There are three essential components in our model. (\textit{i}) We show that the distribution of orientations of bacteria at the funnel gate follows a raised cosine distribution, independent of the funnel geometry (\textit{SI Appendix}, Fig.~\ref{fig:figure2_SI}\textit{E}).  (\textit{ii}) The rule of bacterial alignment with the funnel walls enables us to apply geometrical arguments to determine the fraction of bacteria going through the funnel tip $EF$ after entering the funnel gate. (\textit{iii}) We incorporate the effect of wobbling as a random process following a Gaussian distribution of mean $\mu = 0 \degree$ and standard deviation $\lambda = 30 \degree$ based on independent experimental measurements. With the three components above, our model yields an analytical solution for $N_+(\theta)$, Eq.~\ref{eq:j_up_main},
\begin{figure*}[!ht]
\begin{equation}
    N_{+}(\theta) = \left\{
    \begin{array}{ll}
        \displaystyle \left(2\frac{l}{g}\sin\frac{\theta}{2} + 1\right) \left[ A(\theta) \right] & \displaystyle \forall \; \displaystyle \theta \leq \frac{\pi}{2} \\[10pt]
        \displaystyle \left(2\frac{l}{g}\sin\frac{\theta}{2} + 1\right) \left[A(\theta) + \frac{1}{\pi} \left(1-\cot^2\frac{\theta}{2}\right)  (B(\theta) - \sin B(\theta)\cos(\theta-B(\theta)))\right] & \displaystyle \forall \; \displaystyle  \frac{\pi}{2} < \theta \leq \theta_a  \\[10pt]
        \displaystyle \left(2\frac{l}{g}\sin\frac{\theta}{2} + 1\right) \left[A(\theta) + \frac{1}{\pi} \left(1-\cot^2\frac{\theta}{2}\right)  (B(\theta) - \sin B(\theta)\cos(\theta-B(\theta)))\right] f(\theta) & \displaystyle \forall \; \displaystyle \theta > \theta_a.
    \end{array}
    \right.
\label{eq:j_up_main}
\end{equation}
\end{figure*}
where $A(\theta)=\frac{1}{\pi} (\pi-\theta+\sin\theta)$, $B(\theta) = (\frac{3}{4}+\frac{\sin\theta}{2\pi})(\theta-\frac{\pi}{2})+\frac{\cos\theta}{2\pi}$, and $f(\theta) = \frac{1}{2}\left[ 1 + \text{erf}(\frac{\gamma + \mu}{\sqrt{2} \lambda}) \right]$ with erf denoting the error function (see \textit{SI Appendix}, Sec. II.A for the detailed derivation). Here, $\theta_a$ is a constant depending on the width $w$ and gap $g$ of the funnel wall and $\gamma(\theta)$ is the angle between the orientation of bacteria and the local normal of the walls, which is given analytically as a function of $\theta$ at a given $w$ and $g$ (Fig.~\ref{fig:figure2}\textit{D}, \textit{SI Appendix}, Eq.~\ref{eq:gamma}). 

The term in the first (round) brackets corresponds to the number of bacteria entering the funnel, which increases with $\theta$ across the full range of $\theta$. The term in the second (square) brackets gives the fraction of bacteria that continue to move toward the funnel tip after hitting the interior funnel walls. This term always decreases with $\theta$, despite the different forms at acute and obtuse angles. The competition between the first two terms leads to the non-monotonic trend of $N_+$ and predicts $\theta_{\textrm{max}} = 121.77\degree$, in close agreement with $\theta_{\text{max}} = 120 \degree$ from experiments. The effects of bacterial wobbling and the finite thickness of the funnel walls are captured by the last term $f(\theta)$. This term accounts for the fraction of wobbling bacteria that are rectified after colliding with the opposite wall, a process that is relevant only when $\theta>\theta_a$. Since $\theta_a > \theta_{\textrm{max}}$, the presence of wobbling does not change the prediction of $\theta_{\textrm{max}}$.  

Without fitting parameters, Eq.~\ref{eq:j_up_main} provides a quantitative description of ${N}_{+}(\theta)$ (Figs.~\ref{fig:figure2}\textit{B} and \textit{C}). In the limit of zero wall width ($w \to 0$), both the effect of bacterial wobbling and the interaction with the opposite wall vanish, leading to $f(\theta) \to 1$. The theoretical prediction in this limit agrees with minimal simulations for $\theta < \theta_b$ (Figs.~\ref{fig:figure2}\textit{B} and \textit{C}) and matches minimal simulations with zero wall thickness over the full range of $\theta$ (\textit{SI Appendix}, Fig.~\ref{fig:figure2_SI}\textit{F}). It should be emphasized that, although derived for run-and-tumble particles, our model can be adapted to other types of active particles (e.g. active Brownian particles), when the persistence length of particles is longer than the length of the funnel.

\begin{figure*}[t!]
    \centering
    \includegraphics[width=\linewidth]{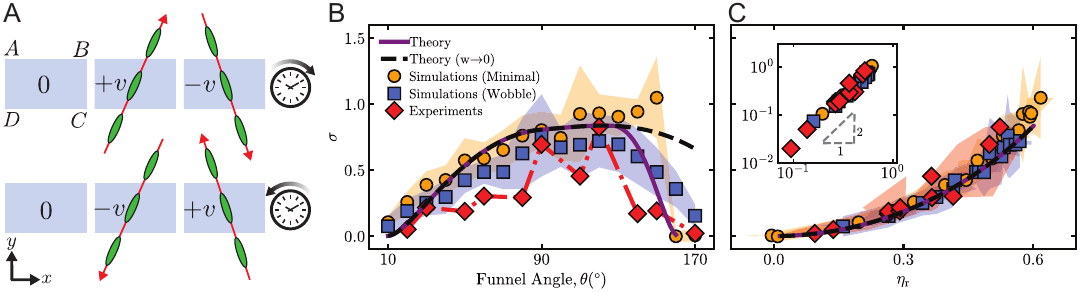}
    \caption{Breaking of local time-reversal symmetry. (\textit{A}) Schematic showing the three discrete states ${0, +v, -v}$ assigned to the region $ABCD$ at the funnel tip. Bacteria are depicted as ellipsoids without flagella to highlight that the orientation of bacteria is not tracked. The lower panels show the time-reversed counterpart of the upper panels. The time-reversed sequence of states is obtained from the time-forward sequence by changing $+v$ to $-v$ and vice versa ($s=0$ states remain invariant under time reversal). Time irreversibility $\sigma$ versus funnel angle $\theta$ ({\textit{B}}) and versus normalized net flux per particle $\eta_r$ (\textit{C}). The inset of (\textit{C}) shows a log-log plot of the same curve in the main plot. Shaded regions denote measurement standard deviations. Experimental $\sigma$ is estimated by concatenating all the time-series at a given $\theta$ in a single sequence (\textit{SI Appendix}, Sec. II.C.1). 
    } 
    \label{fig:figure3}
\end{figure*}

\subsection*{Duality of Two Rectification Systems} How does our system consisting of a single funnel embedded in a uniform bath (System $A$ in Fig.~\ref{fig:figure2}\textit{E}) quantitatively compare with the extensively studied configuration in which an array of funnels separates the bath (viz., microfluidic chamber) into two regions (System $B$ in Fig.~\ref{fig:figure2}\textit{E}) \cite{galajda2007wall, galajda2008funnel, lambert2010collective, kantsler2013ciliary, ro2022model, wan2008rectification, reichhardt2011active, tailleur2009sedimentation}? To answer this question, we first derive a generalized mass transfer relation for bacterial rectification based on our microscopic model (\textit{SI Appendix}, Sec. II.B), which expresses the instantaneous bacterial number flux, $J(t)$, across the funnel tip $EF$ (Fig.~\ref{fig:figure2}A) at time $t$, in terms of two opposing driving forces, 
\begin{equation}
    J(t) = - k \left [\Delta C(t) + \Delta C_{\mathrm{eff}} (t) \right] .
    \label{eq:general_mt}
\end{equation}
Here, $\Delta C(t)$ is the conventional concentration difference between the two regions ($+y$ and $-y$) of the microfluidic chamber, $\Delta C_{\mathrm{eff}} (t)$ is the effective concentration difference capturing the effect of the funnel rectifier, and $k$ is the mass transfer coefficient linking flux to an externally applied concentration difference, which is independent of rectification and can be assessed in a system without a funnel rectifier (\textit{SI Appendix}, Sec. II.B.1, Figs.~\ref{fig:figure3_SI}\textit{A} and \textit{C}).

When applying Eq.~\ref{eq:general_mt} to our system with an isolated funnel (System $A$ in Fig.~\ref{fig:figure2}\textit{E}), the equation reduces to $J^A = -k \Delta C_{\mathrm{eff}}^A$, as bacterial concentration is uniform and thus $\Delta C^A  = 0$. In System $B$ (Fig.~\ref{fig:figure2}\textit{E}), a finite concentration difference is established at steady state with zero steady-state flux $J^B = 0$  \cite{galajda2007wall, galajda2008funnel, lambert2010collective, kantsler2013ciliary, ro2022model, wan2008rectification, reichhardt2011active, tailleur2009sedimentation}. Thus, the steady-state concentration difference across System $B$ is given by $\Delta C^B = -\Delta C_{\mathrm{eff}}^B$. It can be further shown that $\Delta C_{\mathrm{eff}}^A = R\Delta C_{\mathrm{eff}}^B$ where $R = ({N}_{+}+{N}_{-})/2$ (\textit{SI Appendix}, Sec. II.B.3, Eq.~\ref{delta_Ca_delta_Cb}).

Altogether, when applied independently to Systems $A$ and $B$, Eq.~\ref{eq:general_mt} relates the steady-state bacterial flux, $J^A$, produced in System $A$ to the concentration difference, $\Delta C^B$, measured at steady state in System $B$ \cite{galajda2007wall, galajda2008funnel, lambert2010collective, kantsler2013ciliary, ro2022model, wan2008rectification, reichhardt2011active, tailleur2009sedimentation},
\begin{equation}
    J^A = k R \Delta C^B.
    \label{eq:c_a_j_b_main}
\end{equation}
When expressed in terms of the rectification efficiency of the two systems, Eq.~\ref{eq:c_a_j_b_main} yields (\textit{SI Appendix}, Sec. II.B.4) 
\begin{equation}
    \eta_r^A \equiv \frac{N_+^A-N_-^A}{N_+^A+N_-^A} = \frac{\Delta C^B}{2C_0} \equiv \eta_r^B,
    \label{eq:c_a_j_b_eta}
\end{equation}
where $C_0$ is the average bacterial concentration in the whole chamber. Here, we define the rectification efficiency of System $B$, $\eta_r^B$, as concentration difference per particle. The quantitative relation between our system ($A$) and the previously studied system ($B$) reflects the duality of these two processes. Simulations show excellent agreement with the predictions of Eq.~\ref{eq:c_a_j_b_eta} (Fig.~\ref{fig:figure2}\textit{F}).

Finally, using Eq.~\ref{eq:c_a_j_b_eta}, we compare our experimental results ($\eta_r^A$) with those of previous experimental studies ($\eta_r^B$) \cite{galajda2007wall, galajda2008funnel} and find quantitative agreement (Fig.~\ref{fig:figure2}\textit{C}). Notice that previous studies on bacterial rectification fixed the obstacle angle at $\theta = 60\degree$ \cite{galajda2007wall, galajda2008funnel}, which does not correspond to the maximum rectification efficiency, cf. $\theta_{\max} \approx 120\degree$. Furthermore, the generalized mass transfer relation derived here for bacterial rectification (Eq.~\ref{eq:general_mt}) allows us to predict the temporal evolution of concentration difference $\Delta C(t)$ in System $B$ starting from a uniform bacterial bath (\textit{SI Appendix}, Fig.~\ref{fig:figure3_SI}B). The relation can also be applied to quantitatively understand bacterial rectification in systems having more complex geometries beyond System $A$ and $B$ \cite{phan2024social}.  

\section*{Energetics}

\subsection*{Entropy Production} The hallmark of non-equilibrium processes---with rectification in living systems as a particular example---is the presence of non-zero fluxes and, more fundamentally, time-reversal symmetry breaking (TRSB), measured by the entropy production rate (EPR), $\dot{S}$ \cite{jarzynski2011equalities, seifert2012stochastic, peliti2021stochastic}. EPR gives a direct measure of the minimum energetic cost required to maintain a system out of equilibrium and is crucial in understanding the energetics of non-equilibrium processes.

For a system in non-equilibrium steady state (NESS), the Kullback-Leibler divergence (KLD), $\sigma$, between the probability of observing a time-forward trajectory ($\textbf{X}$) and its time-reversed counterpart ($\textbf{X}^\textbf{R}$) bounds $\dot{S}$ as \cite{roldan2014irreversibility, roldan2010estimating, roldan2012entropy, gomez2008footprints, parrondo2009entropy, kawai2007dissipation, roldan2021quantifying, tan2021scale, ro2022model, roldan2010estimating},
\begin{equation}
    \dot{S} \geq \frac{k_B}{\tau} \sigma = \frac{k_B}{\tau} \biggl \langle \ln \frac{P[\textbf{X}]}{P[\textbf{X}^\textbf{R}]} \biggl \rangle,
\label{eq:entropyrate}
\end{equation}
where $k_B$ is the Boltzmann constant, $\tau$ is the sampling interval, and $P[\textbf{X}]$ and $P[\textbf{X}^\textbf{R}]$ are the probability density functions of the time-forward and backward dynamics, respectively. $\langle \cdot \rangle$ denotes average over $P[\textbf{X}]$. The equality in Eq.~\ref{eq:entropyrate} holds asymptotically in the large $\tau$ limit for all mutually absolutely continuous probability measures (viz. $P[\textbf{X}]=0$ if and only if $P[\textbf{X}^\textbf{R}]=0$) \cite{neri2019integral} associated with the degrees of freedom (DOFs) of a system. Even when not all DOFs are measurable, as is often the case in experiments, Eq.~\ref{eq:entropyrate} still provides a lower bound on $\dot{S}$ \cite{cover1991entropy, roldan2021quantifying, roldan2014irreversibility}. More importantly, when applied independently to different DOFs and spatial positions in a system, the ``local'' EPR \cite{ro2022model} quantifies which DOFs and at which spatial positions the dynamics are out of equilibrium.

In the process of bacterial rectification, the time-forward trajectory \textbf{X} can be taken as a time series of the local state $s$ at the tip of the funnel sampled at a constant rate. Specifically, we measure the $y$-component of the position and velocity of bacteria in the tip region \textit{ABCD} (Fig.~\ref{fig:figure3}\textit{A}). Three states are distinguished: $s=0$ when a particle is absent and $s=+v$ $(-v)$ if a particle is present and moving in the $+y$ $(-y)$ direction. The time-reversed dynamics $\textbf{X}^\textbf{R}$ can be obtained by reversing the time series and flipping the sign of $s$, as $s$ is odd under time reversal (Fig.~\ref{fig:figure3}\textit{A}). Note that (\textit{i}) we account only for the direction of velocity, assuming that the magnitude remains constant on average; and (\textit{ii}) we do not track the orientation of bacteria---an even variable under time reversal---and therefore ignore the orientation irreversibility, which is trivially present with or without the funnel rectification because of the self-propulsion of the bacteria \cite{o2022time}. We calculate $\sigma$ using recently introduced KLD estimators based on data compression \cite{ro2022model} (\textit{SI Appendix}, Sec. II.C.1). Figure~\ref{fig:figure3}\textit{B} shows $\sigma$ as a function of the funnel angle $\theta$ from experiments and simulations, exhibiting a non-monotonic trend with a peak around $\theta = \theta_{\max} \approx  120 \degree$.   

As the events of bacteria crossing the funnel tip are independent at low concentrations (\textit{SI Appendix}, Sec. II.C.1),  $P[\textbf{X}]$ and $P[\textbf{X}^\textbf{R}]$ should follow Bernoulli distributions with the probability parameters $p = {N}_{+}/({N}_{+}+{N}_{-})$ and $(1-p) = {N}_{-}/({N}_{+}+{N}_{-})$. The local EPR can then be calculated following Eq.~\ref{eq:entropyrate} as,
\begin{equation}
    \sigma = \frac{{N}_{+} - {N}_{-}}{{N}_{+}+{N}_{-}} \ln\left( \frac{{N}_{+}}{{N}_{-}} \right) = 2 \eta_r^{2} + \frac{2}{3} \eta_r^{4} + \dots,
\label{eq:sigma}
\end{equation}
where in the second equality we Taylor expand the logarithm as $\ln(N_+/N_-) = 2\sum_{n=1}^\infty \eta_r^{2n-1}/({2n-1})$. 

Substituting ${N}_{+}(\theta)$ from the microscopic model (Eq.~\ref{eq:j_up_main}) and assuming constant ${N}_{-}$, we obtain an analytical expression for $\sigma(\theta)$, which shows reasonable agreement with experiments and simulations (Fig.~\ref{fig:figure3}\textit{B}). Since $\sigma$ is a monotonically increasing function of $\eta_r$ (Eq.~\ref{eq:sigma}), it has a maximum corresponding to the maximum in $\eta_r$ at $\theta=\theta_{\max}$ (Fig.~\ref{fig:figure3}\textit{B}). As the rectification efficiency is the normalized bacterial flux, $\eta_r = (N_+-N_-)/(N_++N_-)$, this similarity demonstrates the correlation between the time irreversibility and the local flux of the rectification process, which is a generic feature of non-equilibrium processes.

\begin{figure*}[t!]
    \centering
    \includegraphics[width=\linewidth]{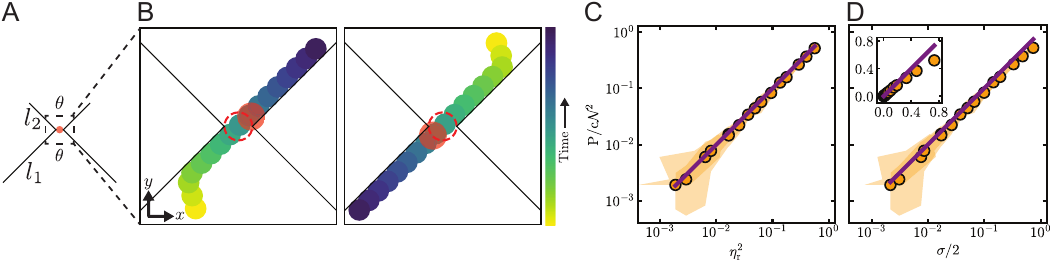}
    \caption{Extractable work from bacterial rectification in an ideal scenario. (\textit{A}) Schematic of a system for measuring extractable work. Two oppositely facing funnel rectifiers with different wall lengths $l_1$ and $l_2$ but the same angle $\theta=90\degree$ are immersed in a bath of bacteria with a harmonically trapped colloid (shown in red) placed at their common tip. (\textit{B}) Superimposed time-series trajectories of bacteria approaching the colloid along the $+y$ (left) and $-y$ (right) directions. The dotted red lines denote the rest position and the filled red circles denote the maximum displacement of the colloid. Comparing normalized extractable power, $P/c\mathcal{N}^2$, with normalized flux per particle, $\eta_r$ (\textit{C}), and time-irreversibility, $\sigma$ (\textit{D}). The solid purple lines are the prediction from Eq.~\ref{eq:power_fl_ep}. The inset of (\textit{D}) shows a linear scale plot of $P/c\mathcal{N}^2$ versus $\sigma/2$. All shaded regions denote measurement standard deviations.
    }
    \label{fig:figure4}
\end{figure*}

Figure~\ref{fig:figure3}\textit{C} shows the measured local EPR as a function of $\eta_r$, verifying the predicted scaling of the local time irreversibility with the local mass (and momentum) flux at the funnel tip (Eq.~\ref{eq:sigma}). Notably, simulations with or without wobble show the same behavior, indicating that the quantitative relation between $\sigma$ and $\eta_r$ is independent of the type of active particles. It is also interesting to notice that the quadratic relation predicted (Eq.~\ref{eq:sigma}) and observed (Figure~\ref{fig:figure3}\textit{C}, inset) for $\eta_r \rightarrow 0$ is the same as the classic result of linear irreversible thermodynamics under the assumptions of linearity between flux and its conjugate thermodynamic force, and of local thermodynamic equilibrium \cite{de2013non}. However, our results assume neither that the system is near equilibrium in the linear regime, nor that the flux is generated by thermodynamic forces. 

Equation~\ref{eq:entropyrate} shows that, when multiplied by $k_BT/\tau$, $\sigma$ estimates the rate at which an irreversible system dissipates heat when connected to a thermal reservoir at temperature $T$ (or equivalently, the energetic cost of maintaining a NESS). Thus, our measurement of $\sigma$ provides a lower bound on the extra energy needed to maintain a steady flux of bacteria ($\sim 300 \, k_B$ s$^{-1}$), in addition to the chemical fuel required for bacterial motility (due to orientation irreversibility present with or without funnel rectification). While it might seem that rectification through \textit{fixed}, passive obstacles has no active energetic cost, a non-zero $\sigma$ clearly demonstrates that energy is indeed required for rectifying bacteria. The extra energy includes but is not limited to the energetic cost of reorientation (flagellar (re)bundling) and the increased dissipative cost of bacterial motion near boundaries compared to the bulk.

\subsection*{Extractable Work} Next, we investigate the amount of work that can be extracted from the process of bacterial rectification. We first examine the relation between fluxes, time irreversibility, and extractable work in an ideal scenario, where an analytical solution can be obtained. A more general case beyond the ideal scenario will be discussed in the next section. In the ideal case, we trap a colloid in a harmonic potential at the common tip of two oppositely-facing funnels with zero wall thickness. The two funnels have the same angle $\theta=90\degree$ but different lengths, $l_1$ and $l_2$ (Fig.~\ref{fig:figure4}\textit{A}). Different bacterial fluxes are generated by varying the ratio $l_1/l_2$. The colloid is weakly coupled to bacterial motion such that the bacterial flux at the funnel tip remains the same with or without the trapped colloid, as quantified by comparing the normalized flux, $\eta_r$, in the two cases (\textit{SI Appendix}, Fig.~\ref{fig:figure5_SI}). Notice that our system is different from autonomous engines studied previously, where asymmetric objects arrest active particles over long periods of time and qualitatively change particle fluxes \cite{angelani2009self, di2010bacterial, pietzonka2019autonomous, angelani2010geometrically, li2013asymmetric}.

Figure~\ref{fig:figure4}\textit{B} shows typical bacterium-colloid interactions in this ideal scenario for bacteria crossing from either the $+y$ or $-y$ direction, where a bacterium pushes the colloid along its self-propulsion direction when passing through the tip. We thus measure the time-averaged $y$-position of the colloid, $y_{c}$, which is non-zero due to a net mass (and momentum) flux in the $+y$ direction. Even though the motion of the colloid is stochastic, rectified bacterial motion provides a non-zero time-averaged driving force $F = Ky_{c}$ against the pulling of the harmonic trap, which allows for work extraction when the force is coupled to an external load. Here, $K$ is the elastic constant of the trap. In the linear regime, the extractable power $P$ is given by $P = F^2/(4\mu)$, where $\mu$ is the mobility of the colloid (\textit{SI Appendix}, Sec. II.C.3). 

\begin{figure*}[t!]
    \centering
    \includegraphics[width=\linewidth]{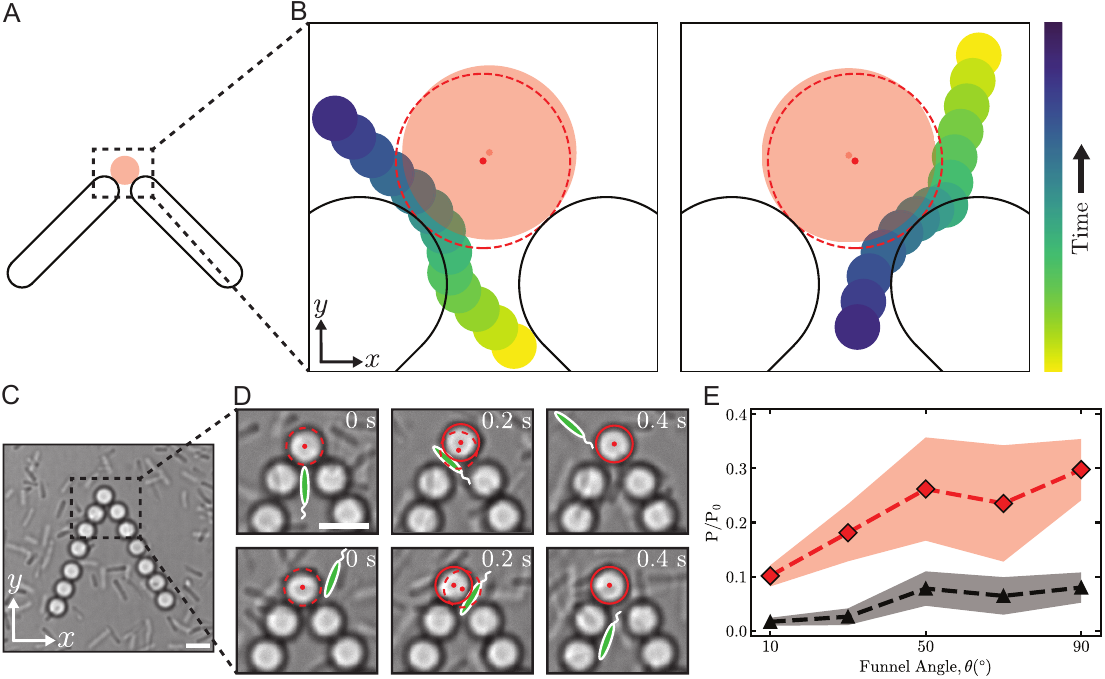}
    \caption{Extractable work from bacterial rectification in a non-ideal scenario. (\textit{A}) Schematic of a system showing a colloid (red) trapped harmonically at the tip of a funnel rectifier. The colloid cannot have large displacements in the $-y$ direction due to the constraint of the wall. (\textit{B}) Superimposed time-series trajectories of bacteria approaching the colloid along the $+y$ (left) and $-y$ (right) directions. The dotted red lines denote the rest position and the filled red circles denote the maximum displacement of the colloid. (\textit{C}) Experimental realization of the system introduced in (\textit{A}), where the funnel ``wall'' is composed of colloids trapped by optical tweezers. The funnel angle is $\theta =50\degree$. The tip colloid is held weakly compared to wall colloids. (\textit{D}) Time series showing typical bacterium-colloid interactions where the tip colloid (the solid red circle) is displaced in the $+y$ direction by the bacteria irrespective of their approaching directions (Movies 3 and 4). Stylized ellipsoids are superimposed on cell bodies for clarity. The positions of the colloid at $t=0$ s are indicated by dashed red lines. Scale bars are 5 $\mu$m. (\textit{E}) Extractable power $P$ as a function of $\theta$. $P$ is normalized by the power of a single bacterium $P_0 = 1.8 \times 10^{-18}$ W (Materials and Methods). Red diamonds are from experiments and black triangles are the prediction of Eq.~\ref{eq:power_fl_ep} in the ideal scenario with experimental fluxes as input. Shaded regions denote measurement standard deviations.
    }
    \label{fig:figure5}
\end{figure*}

From the correlations between $F$, $\eta_r$ and $\sigma$ (\textit{SI Appendix}, Sec. II.C.4), we establish a relation between three fundamental thermodynamic quantities---extractable work ($P$), particle flux ($\eta_r$), and time irreversibility ($\sigma$)---in the non-equilibrium process of bacterial rectification,
\begin{equation}
       P = c\mathcal{N}^2 \eta_r^2 \underset{\eta_r \to 0}{=} \frac{1}{2}c\mathcal{N}^2\sigma,
     \label{eq:power_fl_ep}
\end{equation}
where $\mathcal{N}$ is the total number of bacteria crossing the funnel tip and $c$ is a system-dependent constant (\textit{SI Appendix}, Sec. II.C.4, Eq.~\ref{eq:P_flux}). 

Equation~\ref{eq:power_fl_ep} shows that more power can be harnessed from systems farther away from equilibrium, and that if all the DOFs in a spatial region are locally reversible, $P$, $\eta_r$ and $\sigma$ would all simultaneously be zero \cite{ro2022model}. This interrelationship holds under two general assumptions: (\textit{i}) weak coupling, meaning that $\eta_r$ remains the same with or without the work-extracting mechanism; (\textit{ii}) only one irreversible DOF is coupled to work extraction. Other DOFs are either uncoupled or, if coupled, reversible. Numerical simulations of the ideal scenario show excellent agreement with Eq.~\ref{eq:power_fl_ep} (Figs.~\ref{fig:figure4}\textit{C} and \textit{D}, \textit{SI Appendix}, Sec. I.B).

\subsection*{Beyond Ideal Work Extraction} Is there a way to extract more work from a time-irreversible system than that predicted in the ideal scenario? This can indeed be achieved by relaxing the assumptions in the derivation of Eq.~\ref{eq:power_fl_ep}. Figure~\ref{fig:figure5}\textit{A} shows one possible way to relax the second assumption by a simple modification of the system, where the position of the colloid is chosen such that the funnel walls of finite thickness prevent it from having large displacements in the $-y$ direction (\textit{SI Appendix}, Sec. I.C). In the new scenario, the colloid moves in the $+y$ direction irrespective of the direction of bacterial swimming (Fig.~\ref{fig:figure5}\textit{B}). Thus, the interaction between the colloid and downward-moving bacteria is rectified, which retrieves the lost work in the ideal scenario and increases the total extractable work. While this modified system still obeys the weak coupling assumption, it does not follow the second assumption under which Eq.~\ref{eq:power_fl_ep} was derived. The force exerted by the wall on the tip colloid prevents the colloid from moving freely in the $-y$ direction and therefore acts as an additional irreversible DOF coupled to the colloid.

To demonstrate the possibility of experimentally extracting work from bacterial rectification, we design an experiment that realizes the non-ideal scenario. Specifically, our experiment consists of a funnel-shaped ``wall'' of spherical colloids in a dilute bacterial bath (Fig.~\ref{fig:figure5}\textit{C}). The wall particles are held strongly by the harmonic traps of optical tweezers, whereas the colloid at the tip is loosely held to satisfy the weak coupling condition for work extraction (Materials and Methods). Figure~\ref{fig:figure5}\textit{D} shows typical bacterium-colloid interactions, which are qualitatively similar to the scenario in Fig.~\ref{fig:figure5}\textit{B} (Movies 3 and 4). Our experiments shows clear evidence of extractable work with non-zero $P$, which increases with $\theta$ as more bacteria are rectified at larger $\theta$ (Fig.~\ref{fig:figure5}\textit{E}). Moreover, the power in this non-ideal scenario is consistently higher than that predicted by Eq.~\ref{eq:power_fl_ep}, agreeing with our analysis.

\section*{Discussion}

Bacterial rectification via funnel-shaped obstacles offers an experimentally accessible and analytically tractable model to study rectified living systems. Our results combining experiments, simulations, and theory provide a comprehensive microscopic understanding of the directed transport and energetics of this model system. In particular, we identify the optimal funnel geometry for creating large bacterial flux (or equivalently large concentration difference). We also derive a generalized mass transfer relation for rectified active systems that allows us to extend our microscopic model to previous experimental studies \cite{galajda2007wall, galajda2008funnel} and thus
directly compare them to our results. Further, we quantify the rate of entropy production in this non-equilibrium process, and show that there is an energetic cost associated with bacterial rectification by fixed passive obstacles. Lastly, we devise a weakly-coupled mechanism to probe extractable work from bacterial rectification and uncover a generic relation between local flux, time irreversibility, and extractable work. Altogether, our study sheds light on the fundamental physical principles underlying the rectification of living active matter and facilitates the design of new tools to manipulate the behaviors of active particles for technological applications.

While this study focuses on bacterial rectification, the rectification of microscopically irreversible motion of active particles using passive spatially asymmetric structures has been extensively studied in many other living systems such as \textit{C. reinhardtii} \cite{kantsler2013ciliary}, \textit{S. rosetta} \cite{sparacino2020solitary}, \textit{C. elegans} \cite{nam2013c}, and mammalian sperm cells \cite{guidobaldi2014geometrical}, as well as in systems of non-living active granular particles \cite{galajda2008funnel}. With the knowledge of system-specific particle-boundary interactions, our microscopic model can be modified to yield quantitative predictions for rectification in these systems.

A recent study has shown that carnivorous plants of genus \textit{Genlisea} rectify bacterial motion using their root hairs arranged in the shape of funnels. This anatomical structure creates a flux of soil-dwelling bacteria towards the digestive vesicles of the plants \cite{martin2023carnivorous}. Interestingly, the funnel angle of the root hairs fall in the range $90 \degree - 140 \degree$ \cite{martin2023carnivorous,Carmesin2021plant}, centering around the optimal rectification angle at $\sim 120 \degree$ predicted by our microscopic model and verified by our experiments. This suggests that \textit{Genlisea} could have structurally evolved to achieve optimal rectification, thereby improving their fitness in nutrient-poor soil conditions. 

Our study explores the rectification of bacteria in the dilute limit, where the motions of bacteria are uncorrelated. How do the concepts described here extend to dense systems with correlated bacterial motions? A particularly interesting class of such systems are those displaying active turbulence \cite{liu2021density,alert2022active}. While it has recently been shown that asymmetric obstacles placed in active turbulent baths can move persistently as autonomous engines \cite{sokolov2010swimming, kaiser2014transport} and that the statistics of Lagrangian tracers in active turbulent baths carry information about time-irreversibility \cite{kiran2023irreversibility}, a unifying thermodynamic framework for active turbulence is still missing, which is an interesting topic for future research. 

Is the relation between local fluxes, time irreversibility, and extractable work uncovered in our study applicable to other non-equilibrium systems? Although statistical irreversibility has been widely used to quantify TRSB in various non-equilibrium phenomena \cite{roldan2021quantifying,tan2021scale,ro2022model,muy2013non,tusch2014energy}, simultaneous measurements of time irreversibility, fluxes, and extractable work had not been achieved in any single non-equilibrium systems heretofore. We hope our results will stimulate future research in this direction, which would be of great value for illustrating general principles of non-equilibrium thermodynamics.

\matmethods{
\subsection*{Bacterial Culture} A wild-type \textit{E. coli} strain (BW 25113) genetically modified to express proteorhodopsin (a light-driven transmembrane proton pump) was used in our experiments \cite{peng2021imaging}. The bacterial stock was first inoculated in Terrific Broth (TB) supplemented with ampicillin in an orbital shaker at $35\degree$C for 15 hours. This culture was further diluted (1:100) in fresh TB and grown at $30\degree$C for 6-7 hours, with the addition of 1 mM isopropyl $\beta$-d-1-thiogalactopyranoside and 10 $\mu$M methanolic all-trans-retinal in the mid-log phase to trigger the expression of proteorhodopsin. The bacterial suspension was centrifuged ($800g$ for 5 min) to collect motile cells and the supernatant was resuspended in motility buffer (0.01 M potassium phosphate, 0.067 M NaCl, $10^{-4}$ M EDTA, pH 7.0). The suspension was further washed twice, and the final concentration was measured using a biophotometer via optical density at 600 nm (OD$_{600}$). Bacterial concentrations in all our experiments were fixed below the threshold concentration for collective bacterial swimming \cite{peng2021imaging,liu2021density}. With the supply of oxygen, the average swimming speed of bacteria was 25 $\mu$m/s.

\subsection*{Experiments with PDMS Channels} Polydimethylsiloxane (PDMS) microchannels were fabricated using standard soft lithography. A 4-inch silicon wafer was first immersed in hexamethyldisilazane (HDMS) vapor for 3 min before spin-coating it with a photoresist AZ 1512 (EMD Performance Materials Corp.) at 2500 rpm for 30 s. The wafer was then soft-baked for 1 min at 100 $\degree$C and subsequently exposed to UV light for 5 s through a light field photomask. After a post-exposure bake for 1 min (95 $\degree$C), the wafer was rinsed with DI water and dried. The wafer was then plasma etched to get the final master mold containing a negative replica of the desired channels. The channels were finally cast in PDMS and characterized using optical profilometry (Hyphenated Systems HS200A) (\textit{SI Appendix}, Fig.~\ref{fig:figure6_SI}). Base-washed (NaOH 1M) coverslips and PDMS channels were both soaked with 0.1$\%$ (w$/$v) bovine serum albumin (BSA) in phosphate-buffered saline (PBS) for 1-2 hours to reduce the surface adhesion of bacteria. The channels and coverslips were subsequently washed first with PBS and then twice with DI water (5 min each) before drying and adhering them.

The length, width, and gap of the funnel rectifier were fixed at $30$ $\mu$m, $5$ $\mu$m, and $5$ $\mu$m, respectively. The channel depth was kept at 5-6 $\mu$m, so that bacteria were always close to the focal plane set at the middle of the channels. Strong background flows were often observed in the thin channels due to small variations in the channel height inevitably introduced during fabrication. To combat this adverse effect, a water bridge connecting the inlet and the outlet of the channels was established, which equalizes the hydrostatic pressure difference. Polystyrene colloids of 1 $\mu$m in diameter were introduced as test particles in the channels, which showed Brownian diffusion without any detectable drift, ensuring the effectiveness of the procedure. A bacterial suspension of concentration $8\times10^7$ cells/ml was carefully introduced in the microchannels and imaged via an inverted bright-field microscope (Nikon Ti-Eclipse) using a $40\times$ objective lens with numerical aperture (NA) 0.65. Videos were acquired at a frame rate of 20 frames per second using an Andor Zyla sCMOS camera. Bacteria were manually counted at the funnel tip for each funnel angle.

\subsection*{Experiments with Optical Tweezers} Experiments with optically trapped wall-forming particles (Fig.~\ref{fig:figure5}) were conducted in Hele-Shaw chambers. A small number of 4-$\mu$m-diameter silica particles were mixed with bacterial suspension of concentration $9.6\times10^9$ cells/ml. 5 $\mu$L of the final mixture was deposited on a base-washed glass slide. A glass coverslip was then gently placed over the droplet and the edges were sealed with UV-curable adhesive to reduce background flows. The final Hele-Shaw chamber was quasi-2D with a thickness $\approx 5$ $\mu$m. In the closed chamber, bacteria quickly exhausted the dissolved oxygen. The motility of bacteria was sustained by light with the average swimming speed of bacteria controlled at 12 $\mu$m/s \cite{peng2021imaging}. Colloidal funnel walls were then made by holding the silica particles in optical traps (Aresis Tweez-250si). Trap stiffness was $ K = 3$ pN/$\mu$m for the tip colloid and 5 pN/$\mu$m for the remaining wall colloids. The spacing between colloids was set to 4.2 $\mu$m. An inverted microscope (Nikon Ti-Eclipse) with a $60\times$ objective lens (NA = 1.4) was used for imaging. Videos were obtained at 25 frames per second using an Andor Zyla sCMOS camera. Trajectories of the tip colloid were analyzed using the software Trackpy \cite{allan_daniel_b_2023_7670439}. 

The power of a single bacterium $P_0$ was used to normalize extractable power $P$ (Fig.~\ref{fig:figure5}\textit{E}). Assuming the body of a bacterium as a spheroid of length $l_b$ and width $w_b$ moving in water with viscosity $\eta$, we get \cite{happel1983low}
\begin{equation}
    P_0 = F_{b}v_{b} = 6\pi\eta \frac{w_b}{2} \left[ 1-\frac{1}{5} \left( 1-\frac{l_b}{w_b} \right) \right] v_{b}^2,
    \label{eq:P_0}
\end{equation}
where $F_{b}$ is the thrust force propelling the bacterium and $v_b$ is the velocity of the bacterium. $F_{b}$ is estimated from the drag on the bacterial body due to the force-free feature of the low Reynolds number flow. Taking $l_b = 4$ $\mu$m, $w_b = 0.9$ $\mu$m, $\eta = 8.9\times10^{-4}$ Pa$\cdot$s and $v_{b} = 12$ $\mu$m/s, $P_0 = 1.8 \times 10^{-18}$ W.

\subsection*{Minimal Simulations} Our numerical simulations consisted of $N$ active particles in a 2D box of dimension $L \times L$ with periodic boundary conditions. A funnel rectifier with two rectangular walls of length $l$, width $w$, angle $\theta$, and gap $g$ (\textit{SI Appendix}, Fig.~\ref{fig:figure1_SI}\textit{A}) was placed at the center of the box. Note that the gap $g$ is defined as the distance between the lower corners of the two rectangular funnel walls.  
Due to micro-fabrication limitations, the ends of funnel walls are rounded in our experiments (Figs.~\ref{fig:figure1}\textit{A} and \textit{B}). To incorporate this geometrical feature in our simulations, a round-cornered funnel wall was created based on the rectangular wall by replacing the rectangular end with an inscribed semi-circular arc of diameter $w$ (\textit{SI Appendix}, Fig.~\ref{fig:figure1_SI}\textit{A}).

Bacteria were modeled as point, non-interacting, run-and-tumble particles (RTPs), where the equation of motion for a single particle $i$ in the run phase was given by an overdamped Langevin equation:
\begin{equation}
    \dv{\boldsymbol{r}_i}{t} = v\boldsymbol{\hat{n}}_i + \boldsymbol{\eta}_i
    \label{eq:eqm_1}
\end{equation}
where $\boldsymbol{r}_i$ and $v$ are the position and self-propulsion speed of the particle, respectively. $\boldsymbol{\hat{n}}_i=(\cos\alpha_i,\sin\alpha_i)$ is the unit vector indicating the orientation of the bacterium in the $\alpha_i$ direction with respect to the $x$ axis in the lab frame, and $\boldsymbol{\eta}_i$ is the thermal noise satisfying $\langle \boldsymbol{\eta}_i(t) \rangle = 0$ and $\langle \boldsymbol{\eta}_i(t) \boldsymbol{\eta}_j(t') \rangle = 2D\delta_{ij}\delta(t-t')\mathbb{I}$. Here, $D$ is the diffusion coefficient, and $\mathbb{I}$ is the identity matrix. To mimic a tumbling event, $\alpha_i$ was sampled uniformly from $[0\degree,180\degree)$ after every $m$ simulation time steps. The run-length of the particles is $l_r  = v m \delta t $, where $\delta t$ is the simulation time step. Motivated by experimental observations of the wall-alignment tendency of \textit{E. coli}, particle-wall interactions were modeled using an event-driven approach: if a particle $i$ is set to cross the wall at the next time-step, it is instead placed on the wall and forced to tumble, with the new orientation $\alpha_i$ being chosen such that $\boldsymbol{\hat{n}}(\alpha_i)$ is parallel to the wall according to the rule shown in \textit{SI Appendix}, Fig.~\ref{fig:figure2_SI}\textit{A}.

The length of the funnel rectifier $l$ and the run-time of the active particles $\tau_r \equiv m \delta t$ were chosen as the units of length and time in simulations, respectively. For measurements of fluxes and time-irreversibility shown in Figs.~\ref{fig:figure2} and \ref{fig:figure3}, parameters were set as $L=60$, $N=300$, $l=1$, $w=1/6$, $g=1/6$, $v=30$, $D=0$, $m = 2000$, $\delta t=0.0005$, and $\theta$ was varied from $10\degree-170\degree$. Values of the geometrical parameters of the funnel rectifier ($l, w, g$) were chosen to match the experimental ratios of $l/g = 6$ and $w/g = 1$. Since the active particles are non-interacting, the results are independent of $N$. $L$ was chosen to be larger than the run length of bacteria $l_r = 30$ to minimize finite-size effects. The choice of $v$ and $m$ was such that $l_r \gg l$ to ensure bacterial rectification. All the results are quantitatively the same for any chosen value of $v$ and $\tau$ as long as the criterion $l_r > l$ is satisfied. Thermal noise was neglected with $D=0$, as the random motion of bacteria is dominated by their tumbling.

The application of minimal simulations in different rectification systems (System $A$ and $B$) and systems for extractable work is further explained in \textit{SI Appendix}, Sec. I.

\subsection*{Simulations with Wobbling Particles} Bacterial wobbling is a consequence of the helical run-phase motion of bacteria. An \textit{ab initio} hydrodynamic model of bacterial wobbling is beyond the scope of the current work. Nevertheless, from the perspective of rectification, wobbling essentially adds stochasticity to the bacterium-wall interactions, especially when the rectified bacteria collide with the tip of the opposite wall (Fig.~\ref{fig:figure2}\textit{D}, \textit{SI Appendix}, Fig.~\ref{fig:figure1_SI}). Thus, we model bacterial wobbling phenomenologically by making the otherwise straight-line run-phase motion of the bacteria stochastic.

Specifically, the equation of motion of particles is given by Eq.~\ref{eq:eqm_1}. The tumbling of wobbling particles is implemented using the same method as that of non-wobbling particles: a new orientation $\alpha_i \equiv \alpha_i^*$ is sampled uniformly from $[0\degree,180\degree)$ at the start of every run. However, while for non-wobbling particles, $\alpha_i$ is constant for a given run phase, for a wobbling particle $\alpha_i$ evolves according to an Ornstein-Uhlenbeck process with drift:
\begin{equation}
    \tau_w \dv{\alpha_i}{t} = -\alpha_i + \alpha_i^* + {\zeta}_i
    \label{eq:eqm_wobble}
\end{equation}
where $\tau_w$ is the characteristic time constant for the evolution of $\alpha_i$ and $\zeta_i$ is a Gaussian noise satisfying $\langle \zeta_i(t) \rangle = \mu$ and $\langle \zeta_i(t) \zeta_j(t') \rangle = \lambda^2\delta_{ij}\delta(t-t')$. Here, $\alpha_i^*$ gives the intrinsic orientation of the particle, whereas $\alpha_i$ controls the instantaneous velocity vector of the particle. When the two angles are misaligned, $\alpha_i$ tries to align with $\alpha_i^*$ over a time scale of $\tau_w$. The characteristic length that the particles move before changing their direction due to wobble is $v \tau_w$. 

For wobbling simulations shown in Figs.~\ref{fig:figure2} and \ref{fig:figure3}, parameters were set as $\mu = 0\degree$, $\lambda = 30 \degree$, $\tau_w = 0.005$, $L=10$, $N=50$, $l=1$, $w=1/6$, $g=1/6$, $v=6$, $D=0$, $m = 200$, $\delta t=0.005$, and $\theta$ was varied from $10\degree-170\degree$. The value of $\tau_w$ was chosen to best match experimental data in Fig.~\ref{fig:figure2}\textit{B}. The values of $\mu$ and $\lambda$ were chosen by fitting a Gaussian distribution to the experimental distribution of wobble angles extracted by tracking bacteria swimming in the bulk far from the funnel rectifier.

\subsection*{Theory} The detailed derivation of the microscopic model, the duality of different rectification systems, and the relation between time irreversibility, particle fluxes and extractable work is given in \textit{SI Appendix}, Sec. II.

}

\showmatmethods{} 

\acknow{We thank Emma Jore and Laura Parmeter for help with photolithography, Zhengyang Liu for providing the \textit{E. coli} strain and Buming Guo for providing the compression-based KLD estimator code. We also thank Paul Chaikin, Dipanjan Ghosh, Shashank Kamdar, and Shivang Rawat for fruitful discussions. The research is supported by US NSF CBET 2028652. S.A. and S.M. acknowledge the Simons Center for Computational Physical Chemistry for financial support. Portions of this work were conducted in the Minnesota Nano Center, which is supported by US NSF through the National Nanotechnology Coordinated Infrastructure (NNCI) under Award Number ECCS-2025124. }

\showacknow{} 

\bibsplit[5]

\bibliography{ref}

\end{document}



\title{\textbf{\LARGE{Supplementary Information}} \\Transport and Energetics of Bacterial Rectification}

\author{Satyam Anand}
\email{sa7483@nyu.edu}
\affiliation{Courant Institute of Mathematical Sciences, New York University, New York 10003, USA}
\affiliation{Center for Soft Matter Research, Department of Physics, New York University, New York 10003, USA}
\affiliation{\mbox{Department of Chemical Engineering and Materials Science, University of Minnesota, Minneapolis, MN 55455, USA}}

\author{Xiaolei Ma}
\affiliation{\mbox{Department of Chemical Engineering and Materials Science, University of Minnesota, Minneapolis, MN 55455, USA}}

\author{Shuo Guo}
\affiliation{\mbox{Department of Chemical Engineering and Materials Science, University of Minnesota, Minneapolis, MN 55455, USA}}

\author{Stefano Martiniani}
\email{sm7683@nyu.edu}
\affiliation{Courant Institute of Mathematical Sciences, New York University, New York 10003, USA}
\affiliation{Center for Soft Matter Research, Department of Physics, New York University, New York 10003, USA}
\affiliation{\mbox{Simons Center for Computational Physical Chemistry, Department of Chemistry, New York University, New York 10003, USA}}

\author{Xiang Cheng}
\email{xcheng@umn.edu}
\affiliation{\mbox{Department of Chemical Engineering and Materials Science, University of Minnesota, Minneapolis, MN 55455, USA}}

\maketitle    

\tableofcontents

\section{Simulations}

\subsection{Simulations of Systems $0$, $A$ and $B$}

The equation of motion of particles (Eq.~9) and the interactions between particles and funnel walls were described in the section of Minimal Simulations of Materials and Methods. The interactions between particles and system walls in Systems $0$ and $B$ (solid lines in Fig.~\ref{fig:figure3_SI}\textit{A}) were modeled using a different event-driven approach: if a particle close to a wall crosses it at the next time-step, its orientation is reversed with an updated orientation $\alpha_i^{new} = (\alpha_i^{old}+180\degree)$ mod $360\degree$, such that it reflects back along its direction of approach. The choice of such a reflecting boundary condition, unlike the wall-aligning boundary condition for the funnel walls, reduces undesired effects such as clogging near boundaries. 

For System $0$, the concentration difference between the two sides of the chamber was maintained at all times by removing (adding) a particle at a random position on the $-y$ $(+y)$ side when a particle crossed the funnel gap from $-y$ $(+y)$ to the $+y$ $(-y)$ side. 

For comparison between $J^A$ and $\Delta C^B$ shown in Fig.~2\textit{F}, parameters in both System $A$ and $B$ were set as $L=14$, $N=200$, $l=1$, $w=0$, $g=0.2$, $v=6$, $D=0$, $m = 200$, $\delta t=0.005$, $C_0 = 100/L^{2}$ and $\theta$ was varied from $5\degree-65\degree$. 

For the comparison between $J^0$ and $\Delta C^0$ shown in Fig.~\ref{fig:figure3_SI}\textit{C}, parameters in System $0$ were set as $L=14$, $N=200$, $l=1$, $w=0$, $g=0.2$, $v=6$, $D=0$, $m = 200$, $\delta t=0.005$, $C_0 = 100/L^{2}$ and $\Delta C$ was varied from $20/L^2 - 140/L^2$.

\subsection{Simulations of extractable work in the ideal scenario}

The system for measuring extractable work is the same as that described in minimal simulations with the following modifications. Point-like run-and-tumble particles (RTPs) were replaced by circular disks of diameter $d_p$. A passive, circular colloid of diameter $d_c$ was harmonically trapped at the common tip of two oppositely-facing funnels of zero wall thickness (Fig.~4\textit{A}). The center of the harmonic trap was set to be at the origin (center of the simulation box). The two funnels have the same angle $\theta=90\degree$ but different lengths, $l_1$ and $l_2$. The funnel walls were modeled as hard walls and particle-wall interactions were implemented using an event-driven approach by making two effective virtual funnels of width $(d_p/2)$ and lengths $l_1+(d_p/2)$ and $l_2+(d_p/2)$, and treating particle centers as point particles. Active particles were non-interacting among themselves but interacted with the colloid. The colloid did not interact with the funnel walls. The modified equation of motion of an active particle $i$ was
\begin{equation}
    \dv{\boldsymbol{r}_i}{t} = v\boldsymbol{\hat{n}} + \mu \boldsymbol{F}_{ci} + \boldsymbol{\eta}_i,
    \label{eq:eqm_work}
\end{equation}
where $\mu$ is the mobility and $\boldsymbol{F}_{ci}$ is the force exerted by the colloid on the active particle. The equation of motion of the colloid was
\begin{equation}
    \dv{\boldsymbol{r}_c}{t} =  \mu_c \boldsymbol{F}_{ic} - \mu_c K \boldsymbol{r}_c + \boldsymbol{\eta}_c,
    \label{eq:eqm_colloid}
\end{equation}
where $\boldsymbol{r}_c$ is the position of the colloid, $\mu_c$ is the mobility of the colloid, $K$ is the harmonic trap constant, $\boldsymbol{F}_{ic} = -\boldsymbol{F}_{ci}$ and $\boldsymbol{\eta}_c$ is the thermal noise having the properties $\langle \boldsymbol{\eta}_c(t) \rangle = 0$ and $\langle \boldsymbol{\eta}_c(t) \boldsymbol{\eta}_c(t') \rangle = 2D_c\delta(t-t')\mathbb{I}$. Here, $D_c$ is the diffusion coefficient of the colloid. The interaction force between the active particle and the colloid was modeled as a repulsive spring:
\begin{equation}
    \boldsymbol{F}_{ic} = \epsilon (r_{cut} - r_{ic}) \Theta (r_{cut} - r_{ic}) \hat{\boldsymbol{r}}_{ic} 
    \label{eq:int_force}
\end{equation}
where $r_{ic} = |\boldsymbol{r}_c - \boldsymbol{r}_i|$, $\hat{\boldsymbol{r}}_{ic} = (\boldsymbol{r}_c - \boldsymbol{r}_i)/ r_{ic}$, $\Theta$ is the Heaviside step function, $\epsilon$ controls the strength of the repulsive interaction, and $r_{cut}$ is the cutoff distance for $\boldsymbol{F}_{ic}$.

The mass of the colloid $M_c$ was chosen as a unit of mass for simulations. For the measurement of extractable work shown in Fig.~4, parameters were set as $L=3$, $N=10$, $g=d_c$, $v=1.2$, $D=D_c=0$, $m = 2000$, $\delta t=0.001$, $d_c=0.01$, $d_p=0.007$, $\mu=\mu_c=0.5$, $r_{cut}=(d_c+d_p)/2$, $\epsilon=320$, $k=200$, $\theta=90\degree$, $l_1=1$, and $l_2$ was varied between 0.1 and 1.

\subsection{Simulations of extractable work in the non-ideal scenario}

The simulations for measuring extractable work in the non-ideal scenario consisted of slight modifications of the minimal simulations. Specifically, the center of the colloid was chosen as the intersection point of two lines parallel to the length of the funnel walls and through the middle of the walls. To model the round end of a funnel wall, we placed a semicircle of diameter $w$ at the end of the wall, which interacted with both the colloid and the active particles with an interaction force given by Eq.~\ref{eq:int_force} (with $\epsilon$ and $r_{cut}$ replaced by $\epsilon^w$ and $r_{cut}^w$, respectively). Active particles interacted with the flat part of the funnel wall in the same way as described in the minimal simulations. 

For the modified system depicting non-ideal work extraction (Figs.~5\textit{A} and \textit{B}), parameters were set as $L=10$, $N=10$, $g=0.1$, $v=6$, $D=D_c=0$, $m = 200$, $\delta t=0.005$, $d_c=0.2$, $d_p=0.05$, $M_c = 1$, $\mu=\mu_c=1$, $r_{cut}=(d_c+d_p)/2$, $r_{cut}^w=(d_c+l_2)/2$ for colloid-wall interactions, $r_{cut}^w=(d_p+l_2)/2$ for active particle-wall interactions, $\epsilon=40$, $\epsilon^w=200$, $k=60$, $\theta=90\degree$, $l_1=1$, and $l_2=0.2$.

\begin{figure*}[t!]
    \centering
    \includegraphics[width=1\linewidth]{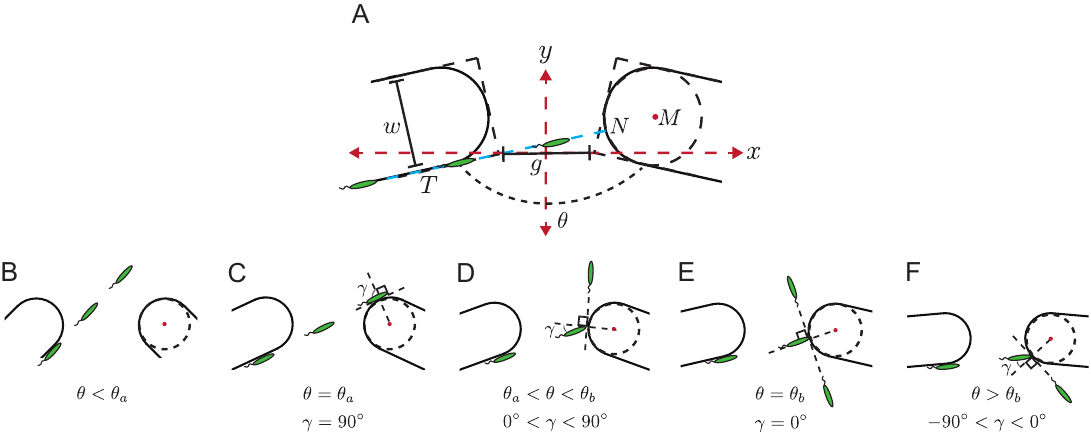}
    \caption{\small Interactions between bacteria and walls near the funnel tip. 
    (\textit{A}) Schematic showing a funnel rectifier with labeled geometric parameters. The construction of the rectifier is described in the section ``Minimal Simulations'' of Materials and Methods. The blue dashed line $T$ indicates the trajectory of a bacterium after its collision with the left wall. The bacterium interacts with the opposite wall on the right before leaving the funnel. Schematics showing the trajectory of non-wobbling bacteria near funnel tip for funnel angle $\theta < \theta_a$ (\textit{B}), $\theta = \theta_a$ (\textit{C}), $\theta_a < \theta < \theta_b$ (\textit{D}), $\theta = \theta_b$ (\textit{E}), and $\theta > \theta_b$ (\textit{F}). $\gamma$ denotes the angle between the orientation of a bacterium and the local normal of the funnel wall at the moment of collision.
    }
    \label{fig:figure1_SI}
\end{figure*}

\section{Theory}

\subsection{Microscopic model}

We develop a microscopic model of bacterial rectification, which predicts the dependence of bacterial fluxes on the geometry of the funnel (Fig.~1\textit{C}). In the following, we first derive a minimal model for the limiting case of zero wall thickness and then include the effect of finite wall thickness and bacterial wobbling to get the full microscopic model. 

\subsubsection{Minimal model} 

In the minimal model, we assume that (\textit{i}) funnel walls are straight lines with zero thickness, $w=0$; (\textit{ii}) the particles are point-like, non-interacting, and undergo run-and-tumble without wobbling; (\textit{iii}) the run length of particles is much longer than the funnel length, $l_r \gg l$; and (\textit{iv}) the gap of the funnel is much smaller than the funnel length, $g \ll l$.

The number flux in 2D is defined as the net flow of the number of particles per unit length and time. The line of interest here is the line $EF$ at the funnel tip (Fig.~2\textit{A}). Since we are only interested in the number of bacteria crossing $EF$ and not the angle of crossing relative to $EF$, the flux we define is a coarse-grained flux. We denote the number of particles crossing $EF$ in the $+y$ $(-y)$ direction per unit length and time as $\mathcal{N}_{+}$ $(\mathcal{N}_{-})$. The net flux through $EF$ can then be written as $(\mathcal{N}_{+} - \mathcal{N}_{-})$. Furthermore, we denote the number of particles moving in $+y$ ($-y$) direction per unit length and time in the bulk far from the funnel as $\mathcal{N}_{+}^{o}$ ($\mathcal{N}_{-}^{o}$).

\begin{figure*}[t!]
    \centering
    \includegraphics[width=1\linewidth]{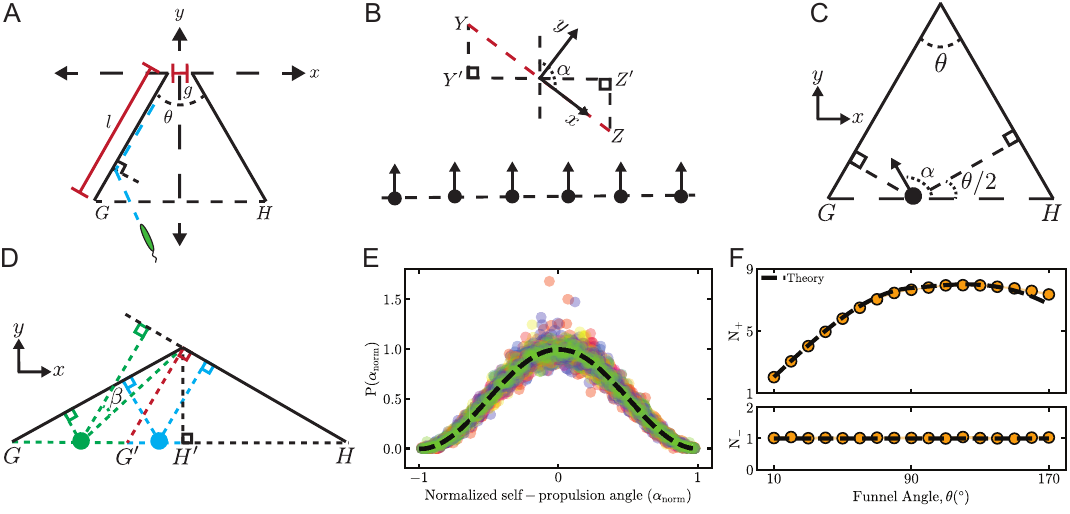}
    \caption{\small Microscopic model of bacterial rectification. 
    (\textit{A}) Schematic showing a funnel rectifier of zero wall thickness. (\textit{B}) Schematic illustrating the origin of the raised cosine distribution of the self-propulsion direction of bacteria $\alpha$ at a line segment. Schematic showing a funnel rectifier with an acute angle (\textit{C}) and an obtuse angle (\textit{D}). (\textit{E}) Probability distribution of the normalized self-propulsion direction ($\alpha_\text{norm}$) reaching the funnel gate $GH$. The dashed black line is the prediction derived from Eq.~\ref{eq:prob_alpha}. Symbols of different colors indicate the results of minimal simulations for funnels of different angles $\theta$. (\textit{F}) The upward and downward fluxes $N_+$ and $N_-$ as a function of $\theta$ for funnels of zero wall thickness. Symbols are from minimal simulations. The dashed black lines show the prediction of $N_+$ from Eq.~\ref{eq:j_up_si} and $N_- = 1$. Shaded regions denote measurement standard deviations. 
    }
    \label{fig:figure2_SI}
\end{figure*}

We start by modeling $\mathcal{N}_{+}$ as the product of (\textit{i}) the number of particles arriving at the funnel opening $GH$ and moving in the $+y$ direction, $\mathcal{N}_{+}^{o}L$, and (\textit{ii}) the fraction of particles that continue to move in the $+y$ direction after a collision with the inside of the funnel walls, $\psi$ (Fig.~\ref{fig:figure2_SI}\textit{A}). Thus, from the conservation of particles, $\mathcal{N}_{+}g = \mathcal{N}_{+}^{o} L \psi$, where $L = 2 l \sin(\theta/2) + g$ is the length of the funnel opening $GH$. As $\mathcal{N}_{+}^{o} \propto \phi v$, where $\phi$ is the number density and $v$ is the self-propulsion speed of particles, we define a normalized particle flux ${N}_{+}$,
\begin{equation}
    {N}_{+} \equiv \frac{\mathcal{N}_{+}}{\mathcal{N}_{+}^{o}} = \frac{L}{g} \psi = \left(2\frac{l}{g}\sin\frac{\theta}{2} + 1\right) \psi,
    \label{eq:norm_n_up}
\end{equation}
which removes the trivial dependency on $\phi$ and $v$. ${N}_{+} = 1$ at any arbitrary line segment chosen in the bulk and at $GH$.

To calculate $\psi$, we first determine the probability distribution of the self-propulsion direction of particles, $\alpha$, crossing a horizontal line segment and moving in the $+y$ direction, $P(\alpha)$. $P(\alpha)$ gives the distribution of the orientation of bacteria arriving at the funnel opening $GH$. Since the motion of particles is random, one may intuitively think that $P(\alpha)$ is a uniform distribution. This, however, is not the case because of two factors. First, consider a case where self-propelled particles uniformly distributed along a line with a line density $n$ move at the same velocity $\mathbf{v}$ upwards (Fig.~\ref{fig:figure2_SI}\textit{B}). A line segment of interest $YZ$ along the $x$ direction with a finite length $\mathcal{L}$ lies above the line and forms an angle $\alpha$ with $\mathbf{v}$ (Fig.~\ref{fig:figure2_SI}\textit{B}). If all particles continue to move in straight lines for a long enough time, $n\mathcal{L}\sin\alpha$ particles would eventually cross $YZ$, as the projection of $YZ$ in the direction normal to $\mathbf{v}$ is $\mathcal{L}\sin\alpha$. This $\sin\alpha$ factor arises from a geometrical reason and is valid when the observation time is long. However, within a finite time interval set by the run time of particles, the number of particles actually reaching $YZ$ is proportional to the magnitude of the normal component of particle velocity with respect to $YZ$, $|\mathbf{v}|\sin\alpha$, giving an additional $\sin\alpha$ factor. The combination of the geometric and kinematic factors dictates that the number of particles reaching $YZ$ is $\sim \sin^2 \alpha$. Noting that $\alpha \in [0, \pi]$, we can write the $P(\alpha)$ as a raised cosine distribution supported on the interval $[\mu_0 - \sigma_0, \mu_0 + \sigma_0]$ with the mean $\mu_0 = \pi/2$ and $\sigma_0 = \pi/2$,
\begin{equation}
    P(\alpha) = \frac{2}{\pi} \sin^2\alpha = \frac{1}{\pi} \left[1 - \cos(2\alpha)\right].
    \label{eq:prob_alpha}
\end{equation}
Equation~\ref{eq:prob_alpha} can be rewritten as a standard raised cosine distribution $P(\alpha_\text{norm})$ having parameters $\mu_0 = 0, \, \sigma_0 = 1$ by normalizing $\alpha$ as $\alpha_\text{norm} = (\alpha - \pi/2) / (\pi/2)$. Figure~\ref{fig:figure2_SI}\textit{E} shows $P(\alpha_\text{norm})$ for varying funnel angles $\theta$ from minimal simulations, which agree well with the standard raised cosine distribution, as predicted by Eq.~\ref{eq:prob_alpha}. 

Going forward, for the $\psi$ calculation, we divide our discussion of the parameter space of funnel angle $\theta \in (0, \pi)$ into two regimes, i.e., acute $\theta$ ($0 < \theta \leq \pi/2$) and obtuse $\theta$ ($\pi/2 < \theta < \pi$). 

\textit{Acute $\theta$}. Once a particle reaches the funnel gate, it goes up ($+y$) after a collision with the inside of funnel walls only if $\alpha \in [\theta/2, \,\, \pi - \theta/2]$ (Fig.~\ref{fig:figure2_SI}\textit{C}). Importantly, this geometric consideration is valid irrespective of the location of the particle at the funnel gate. $\psi$ can then be written as the fraction of $P(\alpha)$ that falls within the range $[\theta/2, \,\, \pi - \theta/2]$,
\begin{equation}
    \psi = 2 \int\limits_{\frac{\pi}{2}}^{\pi - \frac{\theta}{2}} P(\alpha) \,d\alpha	= \frac{1}{\pi} (\pi-\theta+\sin\theta) \equiv A(\theta),
    \label{eq:psi_acute}
\end{equation}
where we have used the fact that $P(\alpha)$ is symmetric around $\alpha = \pi /2$. Combining Eqs.~\ref{eq:norm_n_up} and \ref{eq:psi_acute}, we find $N_+$ for acute angles,
\begin{equation}
    {N}_{+} = \left(2\frac{l}{g}\sin\frac{\theta}{2} + 1\right) A(\theta).
    \label{eq:n_up_acute}
\end{equation}

\textit{Obtuse $\theta$}. We consider only the left half of the funnel gate $GH'$, since the funnel is symmetric around the $y$-axis (Fig.~\ref{fig:figure2_SI}\textit{D}). For a particle landing on $GH'$, the geometric consideration developed previously for acute $\theta$ still applies, which is again independent of the position of the particle. We then further divide  $GH'$ into two segments $GG'$ and $G'H'$ (Fig.~\ref{fig:figure2_SI}\textit{D}), which have lengths $l[\sin(\theta/2) - \cot(\theta/2)\cos(\theta/2)]$ and $l\cot(\theta/2)\cos(\theta/2)$, respectively. For a particle landing on $GG'$, the particle also goes up after a collision if $\alpha \in [\theta/2 - \beta, \,\, \theta/2]$, where $\beta \in [0, \,\, \theta - \pi/2]$ depends on the position of the particle (Figs.~\ref{fig:figure2_SI}\textit{C} and \textit{D}). To estimate this additional fraction of particles $\psi'$ going through the gap from the segment $GG'$, we first find the average of $\beta$, $\beta_\text{avg}$,
\begin{equation}
\begin{split}
    \beta_\text{avg} &= \int\limits_{\pi - \frac{\theta}{2}}^{\frac{\pi}{2} + \frac{\theta}{2}} \beta P(\beta) \,d\beta = \int\limits_{\pi - \frac{\theta}{2}}^{\frac{\pi}{2} + \frac{\theta}{2}} \beta \frac{1}{\pi} (1 - \cos (2\beta)) \,d\beta \\
    &= \left( \frac{3}{4} + \frac{\sin \theta}{2 \pi} \right) \left(\theta - \frac{\pi}{2} \right) + \frac{\cos \theta}{2\pi},
    \label{eq:beta_avg}
\end{split}
\end{equation}
where we use the condition that a particle is equally probable to land anywhere on $GG'$. $\psi'$ is then,
\begin{equation}
    \psi' = \int\limits_{\pi - \frac{\theta}{2}}^{\pi - \frac{\theta}{2} + \beta_{avg}} P(\alpha) \,d\alpha = \int\limits_{\pi - \frac{\theta}{2}}^{\pi - \frac{\theta}{2} + \beta_{avg}} \frac{1}{\pi} [1 - \cos (2\alpha)] \,d\alpha = \frac{1}{\pi} \left[ \beta_{avg} - \sin \beta_{avg} \cos (\theta - \beta_{avg}) \right].
    \label{eq:add_frac}
\end{equation}

Combining Eq.~\ref{eq:add_frac} with the fraction of particles going up from the entire half funnel gap $GH'$ (Eq.~\ref{eq:psi_acute}) and noting that the fractional length of segment $GG'$ compared to $GH'$ is $1-\cot^2(\theta/2)$, we find the total $\psi$ for obtuse $\theta$,
\begin{equation}
    \psi = A(\theta) + \frac{1}{\pi} \left( 1 - \cot^2\frac{\theta}{2} \right) \left[B(\theta) - \sin (B(\theta)) \cos (\theta - B(\theta))\right],
    \label{eq:psi_obtuse_2}
\end{equation}
where we redefine $\beta_\text{ave}$ as $B(\theta)$,
\begin{equation}
    B(\theta) \equiv \beta_\text{avg} = \left( \frac{3}{4} + \frac{\sin \theta}{2 \pi} \right) \left(\theta - \frac{\pi}{2} \right) + \frac{\cos \theta}{2\pi}.
    \label{eq:b_theta}
\end{equation}
Combining Eq.~\ref{eq:psi_obtuse_2} with Eq.~\ref{eq:norm_n_up}, we have ${N}_{+}$ for the obtuse $\theta$ regime as,
\begin{equation}
    {N}_{+} = \left(2\frac{l}{g}\sin\frac{\theta}{2} + 1\right) \left[ A(\theta) + \frac{1}{\pi} \left( 1 - \cot^2\frac{\theta}{2} \right) \left(B(\theta) - \sin (B(\theta)) \cos (\theta - B(\theta))\right) \right]
    \label{eq:n_up_obtuse}.
\end{equation}

Taking Eqs.~\ref{eq:n_up_acute} and \ref{eq:n_up_obtuse} together, we have
\begin{equation}
    N_{+} = \left\{
    \begin{array}{ll}
        \displaystyle \left[2\frac{l}{g}\sin\frac{\theta}{2} + 1\right] [A(\theta)] & \displaystyle \forall \; \displaystyle \theta \leq \frac{\pi}{2} \\[10pt]
        \displaystyle \left[2\frac{l}{g}\sin\frac{\theta}{2} + 1\right] \left[A(\theta) + \frac{1}{\pi} \left(1-\cot^2\frac{\theta}{2}\right)  (B(\theta) - \sin B(\theta)\cos(\theta-B(\theta)))\right] & \displaystyle \forall \; \displaystyle \theta > \frac{\pi}{2}.
    \end{array}
    \right.
\label{eq:j_up_si}
\end{equation}
Since the minimal model has funnel walls of zero thickness, reverse rectification is absent, making $N_- = 1$. The predictions of $N_+$ and $N_-$ quantitatively match the results of minimal simulations with zero wall thickness (Fig.~\ref{fig:figure2_SI}\textit{F}). 

What is the funnel angle corresponding to maximal rectification efficiency, as quantified by $\eta_r (\theta) \equiv ({N}_{+}-{N}_{-})/({N}_{+}+{N}_{-})$? Since $N_{-}$ is constant independent of $\theta$, 
\begin{equation}
    \frac{d\eta_r}{d\theta} = \frac{2N_-}{\left(N_++N_-\right)^2}\frac{dN_+}{d\theta}.
\end{equation}
Because both $N_+$ and $N_-$ are positive, the maximal efficiency is achieved when $dN_+/d\theta = 0$. From Eq.~\ref{eq:j_up_si}, we find $dN_+/d\theta = 0$ when $\theta_{\max} = 121.765 \degree$, in good agreement with experiments ($\theta_{\max} \approx 120 \degree$) (Fig.~2\textit{C}).

\subsubsection{Full model}
We now extend the minimal model to include effects of the finite width of the funnel walls and bacterial wobbling. For funnel angle $\theta < \theta_a \approx 130 \degree$, bacteria do not interact with the opposite funnel wall while passing through the funnel gap (Fig.~\ref{fig:figure1_SI}\textit{B}). The finite wall thickness and bacterial wobbling play only a minor role in this regime. However, at $\theta = \theta_a$, the trajectory of a bacterium becomes tangential to the tip of the opposite wall (Fig.~\ref{fig:figure1_SI}\textit{C}). Above $\theta_a$, a bacterium interacts with the opposite wall before leaving the funnel, where the finite size of the wall and bacterial wobbling strongly affect the rectified trajectory of the bacterium. Here, the key quantity determining the fate of a bacterium is the angle $\gamma$ between the orientation of the bacterium and the local normal of the funnel wall. $\gamma$ is equal to $90 \degree$ at $\theta = \theta_a$ (Fig.~\ref{fig:figure1_SI}\textit{C}), decreases to $0 \degree$ at $\theta = \theta_b$ (Fig.~\ref{fig:figure1_SI}\textit{E}), and becomes negative for $\theta > \theta_b$ (Fig.~\ref{fig:figure1_SI}\textit{F}). 

We first derive an analytical expressions for $\gamma (\theta, w, g)$. The trajectory of a bacterium after collision with one wall of the funnel (Line $T$ in Fig.~\ref{fig:figure1_SI}\textit{A}) passes through the points $(-g/2, 0)$ and $(-g/2 - l\sin{(\theta/2)},- l\cos{(\theta/2}))$. Hence, the equation describing the trajectory is given as,
\begin{equation}
y - x \cot{\left(\frac{\theta}{2} \right)} - \frac{g \cot{\left(\frac{\theta}{2} \right)}}{2} = 0.
\label{eq:line_L}
\end{equation}
We approximate the tip of the funnel wall as a circle centered at $M$ and having a radius $w/2$, given by the equation,
\begin{equation}
\left[- \frac{w \left(- \sin{\left(\frac{\theta}{2} \right)} \cos{\left(\frac{\theta}{2} \right)} - \cos^{2}{\left(\frac{\theta}{2} \right)} + 1\right)}{2 \sin{\left(\frac{\theta}{2} \right)}} + y\right]^{2} + \left[- \frac{g}{2} - \frac{w \left(\sin{\left(\frac{\theta}{2} \right)} + \cos{\left(\frac{\theta}{2} \right)}\right)}{2} + x\right]^{2} - \frac{w^{2}}{4} = 0.
\label{eq:circle_M}
\end{equation}
The bacterial trajectory intersects the circle at the point $N\equiv(N_x,N_y)$ (Fig.~\ref{fig:figure1_SI}\textit{A}) with the coordinates
\begin{align}
N_x = & - \frac{\left(- 2 g \sin{\theta} + 2 g \cot{\left(\frac{\theta}{2} \right)} + \sqrt{2} w \cos{\left(\frac{\theta}{2} + \frac{\pi}{4} \right)} + \sqrt{2} w \cos{\left(\frac{3 \theta}{2} + \frac{\pi}{4} \right)} \right)}{4} \nonumber \\
& + \frac{\left( 2 \sqrt{- 4 g^{2} - 4 \sqrt{2} g w \sin{\left(\frac{\theta}{2} + \frac{\pi}{4} \right)} - w^{2} + \left(2 g \sin{\left(\frac{\theta}{2} \right)} - \sqrt{2} w \cos{\left(\theta + \frac{\pi}{4} \right)}\right)^{2}} \cos{\left(\frac{\theta}{2} \right)}\right) \tan{\left(\frac{\theta}{2} \right)}}{4},  \\
N_y = & \frac{g \sin{\theta}}{2} - \frac{\sqrt{2} w \cos{\left(\frac{\theta}{2} + \frac{\pi}{4} \right)}}{4} - \frac{\sqrt{2} w \cos{\left(\frac{3\theta}{2} + \frac{\pi}{4} \right)}}{4} \nonumber \\
& - \frac{\sqrt{- 4 g^{2} - 4 \sqrt{2} g w \sin{\left(\frac{\theta}{2} + \frac{\pi}{4} \right)} - w^{2} + \left(2 g \sin{\left(\frac{\theta}{2} \right)} - \sqrt{2} w \cos{\left(\theta+ \frac{\pi}{4} \right)}\right)^{2}} \cos{\left(\frac{\theta}{2} \right)}}{2}. \nonumber 
\label{eq:int_N}
\end{align}
$\gamma$ is then obtained by finding the angle between the trajectory (Line $T$) and the local normal line $MN$, given by

\newpage

\begin{sidewaysfigure}
\centering
\begin{equation}
\gamma = \tan^{-1}{\left(\frac{\cot{\left(\frac{\theta}{2} \right)} - \frac{\frac{g \sin{\left(\theta \right)}}{2} - \frac{w \left(- \sin{\left(\frac{\theta}{2} \right)} \cos{\left(\frac{\theta}{2} \right)} - \cos^{2}{\left(\frac{\theta}{2} \right)} + 1\right)}{2 \sin{\left(\frac{\theta}{2} \right)}} - \frac{\sqrt{2} w \cos{\left(\frac{\theta}{2} + \frac{\pi}{4} \right)}}{4} - \frac{\sqrt{2} w \cos{\left(\frac{3\theta}{2} + \frac{\pi}{4} \right)}}{4} - \frac{\sqrt{- 4 g^{2} - 4 \sqrt{2} g w \sin{\left(\frac{\theta}{2} + \frac{\pi}{4} \right)} - w^{2} + \left(2 g \sin{\left(\frac{\theta}{2} \right)} - \sqrt{2} w \cos{\left(\theta + \frac{\pi}{4} \right)}\right)^{2}} \cos{\left(\frac{\theta}{2} \right)}}{2}}{- \frac{g}{2} - \frac{w \left(\sin{\left(\frac{\theta}{2} \right)} + \cos{\left(\frac{\theta}{2} \right)}\right)}{2} - \frac{\left(- 2 g \sin{\left(\theta \right)} + 2 g \cot{\left(\frac{\theta}{2} \right)} + \sqrt{2} w \cos{\left(\frac{\theta}{2} + \frac{\pi}{4} \right)} + \sqrt{2} w \cos{\left(\frac{3\theta}{2} + \frac{\pi}{4} \right)} + 2 \sqrt{- 4 g^{2} - 4 \sqrt{2} g w \sin{\left(\frac{\theta}{2} + \frac{\pi}{4} \right)} - w^{2} + \left(2 g \sin{\left(\frac{\theta}{2} \right)} - \sqrt{2} w \cos{\left(\theta + \frac{\pi}{4} \right)}\right)^{2}} \cos{\left(\frac{\theta}{2} \right)}\right) \tan{\left(\frac{\theta}{2} \right)}}{4}}}{1 + \frac{\left(\frac{g \sin{\left(\theta \right)}}{2} - \frac{w \left(- \sin{\left(\frac{\theta}{2} \right)} \cos{\left(\frac{\theta}{2} \right)} - \cos^{2}{\left(\frac{\theta}{2} \right)} + 1\right)}{2 \sin{\left(\frac{\theta}{2} \right)}} - \frac{\sqrt{2} w \cos{\left(\frac{\theta}{2} + \frac{\pi}{4} \right)}}{4} - \frac{\sqrt{2} w \cos{\left(\frac{3\theta}{2} + \frac{\pi}{4} \right)}}{4} - \frac{\sqrt{- 4 g^{2} - 4 \sqrt{2} g w \sin{\left(\frac{\theta}{2} + \frac{\pi}{4} \right)} - w^{2} + \left(2 g \sin{\left(\frac{\theta}{2} \right)} - \sqrt{2} w \cos{\left(\theta + \frac{\pi}{4} \right)}\right)^{2}} \cos{\left(\frac{\theta}{2} \right)}}{2}\right) \cot{\left(\frac{\theta}{2} \right)}}{- \frac{g}{2} - \frac{w \left(\sin{\left(\frac{\theta}{2} \right)} + \cos{\left(\frac{\theta}{2} \right)}\right)}{2} - \frac{\left(- 2 g \sin{\left(\theta \right)} + 2 g \cot{\left(\frac{\theta}{2} \right)} + \sqrt{2} w \cos{\left(\frac{\theta}{2} + \frac{\pi}{4} \right)} + \sqrt{2} w \cos{\left(\frac{3\theta}{2} + \frac{\pi}{4} \right)} + 2 \sqrt{- 4 g^{2} - 4 \sqrt{2} g w \sin{\left(\frac{\theta}{2} + \frac{\pi}{4} \right)} - w^{2} + \left(2 g \sin{\left(\frac{\theta}{2} \right)} - \sqrt{2} w \cos{\left(\theta + \frac{\pi}{4} \right)}\right)^{2}} \cos{\left(\frac{\theta}{2} \right)}\right) \tan{\left(\frac{\theta}{2} \right)}}{4}}} \right)}.
\label{eq:gamma}
\end{equation}
\end{sidewaysfigure}

\cleardoublepage

At $\theta = \theta_a$, Line $T$ is tangent to the circle centered at $M$. By setting the distance between Line $T$ and Point $M$ to be equal to the radius of the circle $w/2$, we get $\theta_a = 129.219 \degree$ for our funnel geometry with $w = g$. At $\theta = \theta_b$, Line $T$ passes through the center $M$ of the circle. By setting the distance between Line $T$ and Point $M$ to be equal to $0$, we get $\theta_b = 153.739 \degree$ for $w = g$.

With the analytical solution of $\gamma$, we are ready to incorporate the effect of the finite wall thickness and bacterial wobbling in our microscopic model. To make the problem analytically tractable, we assume that after getting rectified from one side of the funnel wall, bacteria continue moving along a straight-line path until they hit the circular end of the opposite funnel wall (Figs.~\ref{fig:figure1_SI}\textit{C}-\textit{F}). Without wobbling, $\gamma$ alone dictates whether a bacterium goes in the $+y$ or $-y$ direction after collision with the funnel wall. For $\gamma > 0$, the bacterium goes in the $+y$ direction after interacting with the opposite funnel wall, whereas for $\gamma < 0$, the bacterium goes in the $-y$ direction. With bacterial wobbling, an effective $\gamma_{\text{eff}}$ is given as $\gamma + \kappa$, where $\kappa$ is the deviation of the bacterial body orientation from its self-propulsion direction due to wobbling when it reaches the intersection point $N$. We measure the experimental distribution $\kappa$ of wobble angles by tracking bacteria swimming in the bulk far from the funnel rectifier and fit the distribution with a Gaussian distribution of a mean $\mu = 0 \degree$ and a standard deviation $\lambda = 30 \degree$. For a bacterium to go in the $+y$ direction after interacting with the opposite funnel wall, $\gamma_{\text{eff}} = \gamma + \kappa$ should be $>0$. Thus, among all the bacteria arriving at the funnel gap, the fraction of bacteria $f$ that continue to go up after interacting with the opposite wall is given as,
\begin{equation}
f =\text{Prob.}(\gamma_{\text{eff}} > 0) = \text{Prob.}(\kappa > -\gamma) = 1 - \text{CDF}(-\kappa) = 1 - \frac{1}{2}\left[ 1 + \text{erf}\left(\frac{-\gamma - \mu}{\sqrt{2} \lambda}\right) \right] = \frac{1}{2}\left[ 1 + \text{erf}\left(\frac{\gamma + \mu}{\sqrt{2} \lambda}\right) \right],
\label{eq:wobble_frac}
\end{equation}
where CDF denoted the cumulative density function of the Gaussian distribution of $\kappa$, and erf denotes the error function given by $\text{erf}(z) = (2/\sqrt\pi) \int_{0}^{z} e^{-t^2} dt$. Substituting $\gamma$ from Eq.~\ref{eq:gamma} into Eq.~\ref{eq:wobble_frac}, we get $f (\theta, w, g)$. As $\theta_a > 90\degree$, we only need to modify the solution of $N_+$ (Eq.~\ref{eq:j_up_si}) in the regime of obtuse angles, 
\begin{equation}
    N_{+} =  \left[2\frac{l}{g}\sin\frac{\theta}{2} + 1\right] \left[A(\theta) + \frac{1}{\pi} \left(1-\cot^2\frac{\theta}{2}\right)  (B(\theta) - \sin B(\theta)\cos(\theta-B(\theta)))\right] f \; \; \; \; \theta \geq \theta_a.
   \label{eq:j_up_si_wobble}
\end{equation}
In combination with Eq.~\ref{eq:j_up_si}, we reach the final expression of $N_+$ given by Eq.~1 of the main text.

\subsection{Generalized mass transfer relation}

A system produces a mass flux $J$ in response to an applied thermodynamic force, i.e., a concentration difference $\Delta C$. There exists a generic relation, the so-called mass transfer relation, $J = - k \Delta C$, where $k$ is the mass transfer coefficient with a unit of length/time \cite{cussler2009diffusion}. The relation, typically valid for small $\Delta C$, is the mass-transfer analog to Newton's law of cooling for heat transfer. For non-interacting run-and-tumble particles (RTPs), the linear relation $J = - k \Delta C$ holds even at high $\Delta C$. However, the introduction of a funnel rectifier modifies the motion of particles, generating a non-zero $J$ even in the absence of $\Delta C$. Here, we derive a generalized mass transfer relation for bacterial rectification, which allows for a quantitative prediction of several important features of the process.

Our discussion of the mass transfer relation is divided into three parts. First, we obtain $k$ in a simple slit geometry without rectification (System $0$, the left schematic in Fig.~\ref{fig:figure3_SI}\textit{A}). We then use the microscopic model described above to reformulate the funnel rectification in terms of an effective density difference $\Delta C_{\mathrm{eff}}$ (System $A$, the middle schematic in Fig.~\ref{fig:figure3_SI}\textit{A}). Finally, using $\Delta C_{\mathrm{eff}}$, we derive a mass transfer relation for bacterial rectification in the extensively-studied geometry \cite{drocco2012bidirectional, martinez2020trapping, guidobaldi2014geometrical, wan2008rectification, reichhardt2011active, tailleur2009sedimentation, galajda2007wall, galajda2008funnel, lambert2010collective, ro2022model, kantsler2013ciliary, sparacino2020solitary, nam2013c, guidobaldi2014geometrical, galajda2008funnel}, where a funnel rectifier embedded in a partition separating bacterial bath into two regions (System $B$, the right schematic in Fig.~\ref{fig:figure3_SI}\textit{A}). The average number concentration is fixed to be $C_0$ in all three geometries. The total area of the chamber is $2 V$, making the area of either side ($+y$ or $-y$) of the chamber $V$ in Systems $0$ and $B$. Total number of particles in the chamber, $2C_0V$, is conserved in all three geometries.

\subsubsection{System $0$}

\begin{figure*}[t!]
    \centering
    \includegraphics[width=\linewidth]{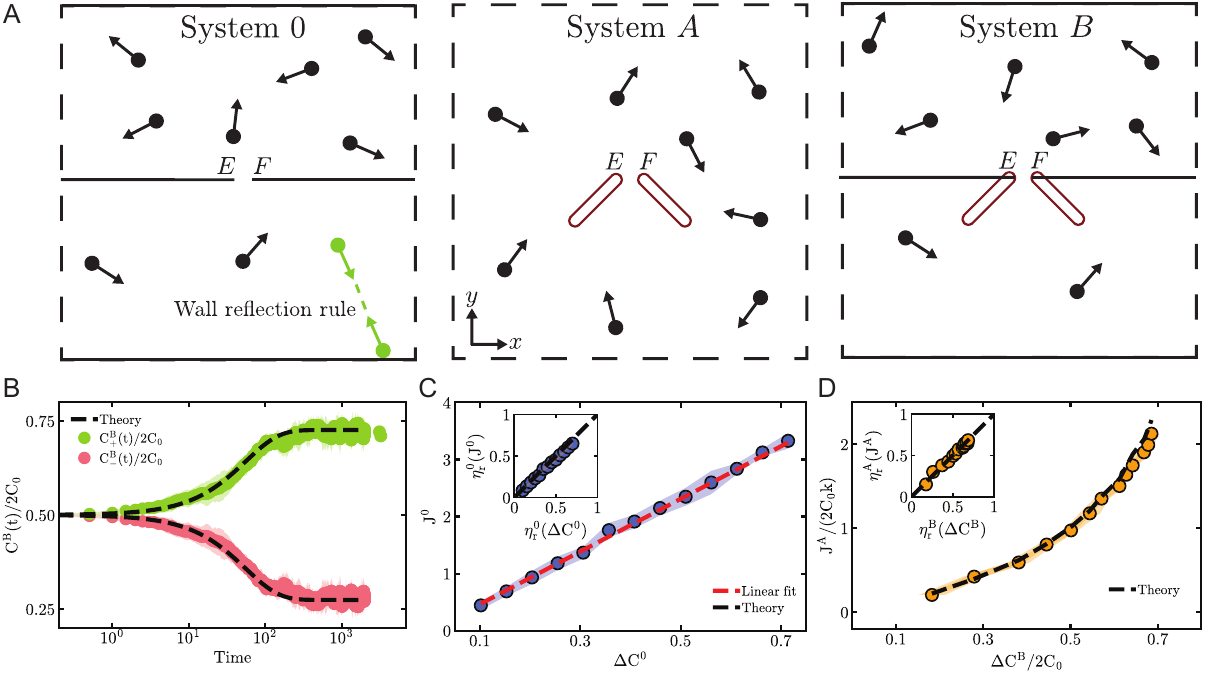}
    \caption{\small Mass transfer relation for bacterial rectification. (\textit{A}) Schematics showing three different Systems: $0$, $A$, and $B$. Funnel rectifiers are shown in red. Solid black lines represent solid boundary walls with a reflecting boundary condition and dashed black lines denote the locations with periodic boundary conditions. Flux is measured in all three systems at the imaginary line $EF$ at the gap connecting the $+y$ and $-y$ sides of the chamber. (\textit{B}) Time evolution of concentration in $+y$ and $-y$ regions in System $B$ starting from an initially uniform concentration of $C_0$ for funnel angle $\theta = 20\degree$. Dashed black lines are predictions of Eq.~\ref{eq:C_t}. (\textit{C}) Particle flux $J$ as a function of the concentration difference $\Delta C$ measured in System $0$. The red dashed line shows a linear fit used to extract the mass transfer coefficient $k$. Inset: Normalized flux per particle $\eta_r^0 (J)$ plotted against normalized concentration difference per particle $\eta^0_r (\Delta C)$ for System $0$. (\textit{D}) $J$ measured in System $A$ as a function of $\Delta C$ measured in System $B$. The dashed line is the prediction of Eq.~\ref{eq:c_a_j_b}. Inset: Normalized flux per particle in System $A$, $\eta^A_r$, plotted against normalized concentration difference per particle in System $B$, $\eta^B_r$. Black dashed lines in (\textit{C}) and (\textit{D}) are the prediction of Eq.~\ref{eq:j_norm_c_norm}. All shaded regions denote measurement standard deviations.
    }
    \label{fig:figure3_SI}
\end{figure*}

We denote the number concentration of the region located on the $+y$ ($-y$) side of System $0$ as $C_{+}^0$ ($C_{-}^0$) (Fig.~\ref{fig:figure3_SI}\textit{A}). The concentration difference $\Delta C^0$ between the $+y$ and $-y$ regions of the chamber is then
\begin{equation}
\Delta C^0 = C_{+}^0 - C_{-}^0.
\label{eq:density_difference}
\end{equation}
We define $\mathcal{N}_{+}^0$ $(\mathcal{N}_{-}^0)$ as the number of particles crossing the gap $EF$ in the $+y$ $(-y)$ direction per unit length and time. Thus, the coarse-grained number flux at $EF$ is simply $J^0 = \mathcal{N}_{+}^0 - \mathcal{N}_{-}^0$. We enforce a time-independent constant concentration difference, $\Delta C^0$, between the two sides of the chamber in our study. Such a density difference sustains a finite flux $J$. It can easily be seen that in System $0$, the number of particles crossing $EF$ in the $+y$ and $-y$ direction per unit length and time are  $\mathcal{N}_{+}^0 = k C_{-}^0$ and $\mathcal{N}_{-}^0 = k C_{+}^0$, respectively. As a result, $J^0 = -k \Delta C^0$. Note that the number of particles crossing $EF$ in the $+y$ direction, $\mathcal{N}_{+}^0$, depends on the particle concentration in the $-y$ region $C_{-}^0$, and vice versa. System $0$ remains at a steady state at all times. 

We use System $0$ to measure the mass transfer coefficient $k$ in our minimal simulations. Specifically, we impose different control concentration differences, $\Delta C^0$, across the slit in System $0$, and measure the corresponding $J^0$. $k$ is then determined directly via a linear fit of $J^0$ versus $\Delta C^0$ (Fig.~\ref{fig:figure3_SI}\textit{C}). $k$ depends on the velocity and type of active particles, both of which are fixed in our simulations.

\subsubsection{System $A$}

Next, we consider the effect of funnel rectification on the mass transfer in System $A$ (Fig.~\ref{fig:figure3_SI}\textit{A}). Periodic boundary conditions are applied on all four boundaries of the simulation box and there is no boundary separating the $+y$ and $-y$ regions of the chamber. Thus, the concentration difference between the $+y$ and $-y$ regions of the chamber is zero, viz., $C_{+}^A = C_{-}^A = C_0$ and $\Delta C^A = 0$ in System $A$. System $A$ also remains at a steady state at all times. 

The flux $J$ follows the same definition as System $0$, i.e., $J^A = \mathcal{N}_{+}^A - \mathcal{N}_{-}^A$. The funnel rectifier enhances the mass transfer along the $+y$ and $-y$ directions, which are quantified by the normalized flux $N_{+}$ and $N_{-}$. $N_{+}$ depends on the funnel geometry and is given by Eq.~\ref{eq:j_up_si} and $N_{-} = 1$. Here, we consider only the limiting case of zero wall thickness. Thus, we can formally write $\mathcal{N}_{+}^A = k N_{+} C_{-}^A = k N_{+} C_{0}$ and $\mathcal{N}_{-}^A = k N_{-} C_{+}^A = k N_{-} C_{0}$, where $k$ is the aforementioned mass transfer coefficient. Hence, $J = k C_0 (N_{+} - N_{-})$.

Since $\Delta C^A = 0$, the driving force leading to a finite $J^A$ must be solely due to rectification. To map the thermodynamic force due to rectification onto an effective concentration difference, $\Delta C^A_{\mathrm{eff}}$, $J^A$ can be rewritten as $J^A = - k C_0 (N_{-} - N_{+}) \equiv - k \Delta C^A_{\mathrm{eff}}$. Here, we define an effective concentration difference in System $A$ as  
\begin{equation}
    \Delta C^A_{\mathrm{eff}} \equiv C_0 (N_{-} - N_{+}),
    \label{eq:effective_density_difference_A}
\end{equation}
which is the driving force responsible for non-zero $J^A$ in the presence of a funnel rectifier.

\subsubsection{System $B$}

With the mass transfer coefficient $k$ determined in System $0$ and the effective density difference $\Delta C_{\mathrm{eff}}$ defined in System $A$, we are ready to derive a generalized mass transfer relation for bacterial rectification in System $B$, a geometry extensively studied in previous works \cite{drocco2012bidirectional, martinez2020trapping, guidobaldi2014geometrical, wan2008rectification, reichhardt2011active, tailleur2009sedimentation, galajda2007wall, galajda2008funnel, lambert2010collective, ro2022model, kantsler2013ciliary, sparacino2020solitary, nam2013c, guidobaldi2014geometrical, galajda2008funnel}. Specifically, we denote the number concentration of particles in regions located on the $+y$ ($-y$) side of System $B$ at time $t$ as $C_{+}^B(t)$ ($C_{-}^B(t)$). The concentration difference between the $+y$ and $-y$ regions of the chamber is then $\Delta C^B(t) = C_{+}^B(t) - C_{-}^B(t)$. The coarse-grained number flux at $EF$ is simply $J^B(t) = \mathcal{N}_{+}^B(t) - \mathcal{N}_{-}^B(t)$, following the analogous definition from System $0$. Note that all the quantities defined in System $B$ are functions of time. The steady state is reached when $t \to \infty$. 

We start with the analysis of the steady-state density difference and note that $J^B(t \to \infty) = \mathcal{N}_{+}^B(t \to \infty) - \mathcal{N}_{-}^B(t \to \infty) = 0$. Using the analogous definition from System $A$, $\mathcal{N}_{+}^B = k N_{+} C_{-}^B$ and $\mathcal{N}_{-}^B = k N_{-} C_{+}^B$. At steady state, $J^B = 0$, which requires $\mathcal{N}_{+}^B = \mathcal{N}_{-}^B$. Thus, we have $N_{+} / N_{-} = C_{+}^B / C_{-}^B$, which, when combined with $C_{+}^B + C_{-}^B = 2C_0$, gives 
\begin{equation}
\Delta C^B = 2 C_0 \frac{N_+-N_-}{N_++N_-} \equiv - \Delta C_{\mathrm{eff}}^B.
\label{eq:effective_concentration_difference_B}
\end{equation}
In comparison with Eq.~\ref{eq:effective_density_difference_A}, we see that the effective concentration differences in System $A$ and $B$ are related via  
\begin{equation}
    \Delta C_{\mathrm{eff}}^A = \frac{N_++N_-}{2}\Delta C_{\mathrm{eff}}^B \equiv R \Delta C_{\mathrm{eff}}^B, 
\label{delta_Ca_delta_Cb}
\end{equation}
a relation that is used in the main text.

We now move on to the analysis of the temporal evolution of $C_{+}^B(t)$ and $C_{-}^B(t)$ starting from an initial concentration $C_0$ on both $+y$ and $-y$ sides of the chamber. Since $\mathcal{N}_{+}^B(t) = k N_{+} C_{-}^B(t)$ and $\mathcal{N}_{-}^B(t) = k N_{-} C_{+}^B(t)$, 
\begin{equation}
J^B(t) = \mathcal{N}_{+}^B(t) - \mathcal{N}_{-}^B(t) = k \left(N_{+} C_{-}^B(t) -  N_{-} C_{+}^B(t)\right).    
\label{eq:J}
\end{equation} 
Two competing driving forces dictate the flux $J^B(t)$, i.e., the concentration difference $\Delta C^B(t)$, and the presence of the funnel rectifier captured by $\Delta C_\mathrm{eff}^B(t)$. To decompose their respective contributions, we rewrite $J^B(t)$ as,
\begin{widetext}
\begin{equation}
\begin{split}
    J^B(t) & = - k \left[ \Delta C^B(t) \right] - k \left[(N_{-} - 1)C_{+}^B(t) - (N_{+} - 1)C_{-}^B(t)\right] \\
    & = - k \left[ \Delta C^B(t) \right] - k \left[ \Delta C^B_{\mathrm{eff}}(t) \right], 
    \label{eq:delta_c_rect}
\end{split}
\end{equation}
\end{widetext}
where $\Delta C^B_{\mathrm{eff}}(t) \equiv (N_{-} - 1)C_{+}^B(t) - (N_{+} - 1)C_{-}^B(t)$. Equation~\ref{eq:delta_c_rect} is the generalized mass transfer relation connecting the flux to the concentrations on the two sides of the chamber. The first term denotes the contribution to the number flux due to the actual concentration difference $\Delta C^B(t)$. The second term, arising from the presence of the funnel rectifier, is the rectification force expressed in terms of an effective concentration difference $\Delta C^B_{\mathrm{eff}}(t)$. At $t = 0$ with $C_{-}^B(t) = C_{+}^B(t) = C_0$, $\Delta C ^B_{\mathrm{eff}}(t=0) = C_0 (N_--N_+)$, which is the same as $\Delta C^A_{\mathrm{eff}}$ (Eq.~\ref{eq:effective_density_difference_A}) as expected. When $t \to \infty$ in steady state, $C_+^B/C_-^B = N_+/N_-$. Thus, $\Delta C^B_{\mathrm{eff}}(t\to\infty) = 2C_0(N_--N_+)/(N_-+N_+)$, a relation that has been shown in Eq.~\ref{eq:effective_concentration_difference_B}.  In the absence of a rectifier, $N_{+} = N_{-} = 1$ and $\Delta C^B_{\mathrm{eff}}(t) = 0$, which restores to the result of System 0. 

With Eq.~\ref{eq:delta_c_rect}, the temporal evolution of $C_{+}^B(t)$ can be written as,
\begin{equation}
    \dv{C_{+}^B(t)}{t} = \frac{g}{V} J^B(t) = k \frac{g}{V} \left[ N_{+} C_{-}^B(t) - N_{-} C_{+}^B(t) \right],
    \label{eq:dC_up}
\end{equation}
where $g$ is the length of gap $EF$ and $2V$ is the total area of the chamber.

Solving Eq.~\ref{eq:dC_up} with the initial condition $C_{+}^B(t) = C_0$ at $t = 0$ and using the fact that $C_{+}^B(t) + C_{-}^B(t) = 2C_0$, we have,
\begin{equation}
\begin{aligned}
      C_{+}^B(t) = C_0 \left[ \frac{2N_+}{N_++N_-} - \left( \frac{N_+-N_-}{N_++N_-} \right) \exp\left(-\frac{kgN_{+}+kgN_{-}}{V} t\right)  \right] \\
      C_{-}^B(t) = C_0 \left[ \frac{2N_-}{N_++N_-} + \left( \frac{N_+-N_-}{N_++N_-} \right) \exp\left(-\frac{kgN_{+}+kgN_{-}}{V} t\right)  \right].
    \label{eq:C_t}
\end{aligned}
\end{equation}
$J^B(t)$ and $\Delta C^B(t)$ then follow from their respective definitions (Eq.~\ref{eq:J}). Figure~\ref{fig:figure3_SI}\textit{B} shows a comparison between the prediction of Eq.~\ref{eq:C_t} and the results of minimal simulations, which shows an excellent agreement.

\subsubsection{Duality between Systems $A$ and $B$}

The concentration difference in System $B$, $\Delta C^B$, can be related to the number flux at the funnel tip in System $A$, $J^A$. Using the fact that $\Delta C^B = 2 C_0 (N_+-N_-)/(N_++N_-)$ and $J^A = k C_0 (N_{+} - N_{-})$, we have
\begin{equation}
    J^{A} = k \left( \frac{N_{+} + N_{-}}{2}\right) \Delta C^B = k R \Delta C^B.
    \label{eq:c_a_j_b}
\end{equation}
This is the result shown in Eq.~3 in the main text. Here, $k R$ is an effective mass transfer coefficient and $R \equiv (N_{+} + N_{-})/2$ is a correction factor for rectification. Note that Eq.~\ref{eq:c_a_j_b} is in the form of a mass transfer relation, where $J^{A}$ is a number flux and $\Delta C^B$ is an applied concentration difference. The crucial difference is that $J^{A}$ and $\Delta C^B$ correspond to two different systems, but can nevertheless be directly related, revealing a correspondence between the two systems. Figure~\ref{fig:figure3_SI}\textit{D} shows that numerically measured $J^A$ and $\Delta C^B$ are in good agreement with the prediction of Eq.~\ref{eq:c_a_j_b}. 

While $R$ is predicted by the microscopic model of rectification (Eq.~\ref{eq:j_up_si}), when the funnel geometry is complicated, it may not be possible to write $R$ analytically. The quantity has to be measured either numerically or in experiments. In fact, Eq.~\ref{eq:c_a_j_b} can be recast in a normalized form to remove the dependency on any parameters as,
\begin{equation}
    \eta^{A}_r \equiv \frac{\mathcal{N}_{+}^{A} - \mathcal{N}_{-}^{A}}{\mathcal{N}_{+}^{A} + \mathcal{N}_{-}^{A}} = \frac{N_+-N_-}{N_++N_-} = \frac{\Delta C^B}{2C_0} \equiv \eta^{B}_r,
    \label{eq:j_norm_c_norm}
\end{equation}
where $\eta^{A}_r$ is the normalized number flux per particle in System $A$, $\eta^{B}_r$ is the normalized concentration difference per particle in System $B$. In deriving the relation, Eq.~\ref{eq:effective_concentration_difference_B} is used to relate the flux to the concentration difference. Note that Eq.~\ref{eq:j_norm_c_norm} is valid irrespective of whether $N_+$ and $N_-$ can be solved analytically. The prediction of Eq.~\ref{eq:j_norm_c_norm} compares well with the result of minimal simulations (Fig.~2\textit{F} and Fig.~\ref{fig:figure3_SI}\textit{D} inset). Finally, the relation between the imposed concentration difference $\Delta C^0$ and the resulting flux $J^0$ in System $0$ can also be expressed in a similar fashion, giving $\eta^{0}_r (J^0) = \eta^0_r (\Delta C^0)$, where $\eta^0_r(J^0)$ and $\eta^0_r (\Delta C^0)$ are the normalized flux and the normalized concentration difference in System 0, respectively, in good agreement with numerical simulations (Fig.~\ref{fig:figure3_SI}\textit{C} inset).

\subsection{Time-irreversibility, fluxes, and extractable work}

\subsubsection{Characterization of time-irreversibility} 

Characterization of time-irreversibility is done by measuring the Kullback-Leibler divergence (KLD), $\sigma$, between two probability distributions: the time-forward distribution $P[\textbf{X}]$ and the time-reversed distribution $P[\textbf{X}^\textbf{R}]$ of coarse-grained states. The local state $s$ of the region $ABCD$ close to the funnel tip can take three different values at any time: $s=0$ when no bacteria are present, and $s=+v$ $(-v)$ if a bacterium is present and moving in the $+y$ $(-y)$ direction (Fig.~3\textit{A}). From the time-forward sequence of states $\textbf{X}$, the corresponding time-reversed sequence $\textbf{X}^\textbf{R}$ is obtained by changing $+v$ to $-v$ and vice versa. $s=0$ remain unchanged in the time reversal. A typical time-forward trajectory of states takes the form, $\textbf{X} = (+v,0,+v,...,0,0,-v,0)$, and its time-reversed counterpart is given by, $\textbf{X}^\textbf{R} = (0,+v,0,0,...,-v,0,-v)$. The $s=0$ state can be filtered out since it remains the same during the time-reversal operation and crossing events are independent of one another. The procedure gives $\textbf{X} = (+v,+v,...,-v)$ and $\textbf{X}^\textbf{R} = (+v,...,-v,-v)$ for the previous example.

\begin{figure*}[t!]
    \centering
    \includegraphics[width=\linewidth]{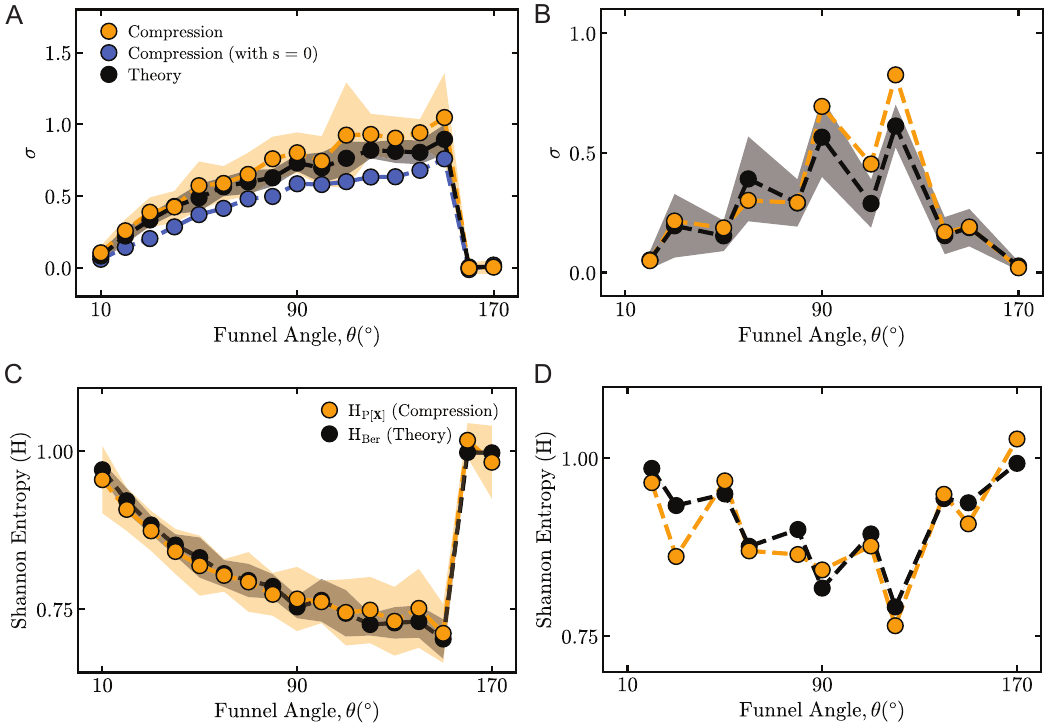}
    \caption{\small Time-irreversibility and Shannon entropy measured using compression-based estimators. Time-irreversibility $\sigma$ as a function of the funnel angle $\theta$ from minimal simulations (\textit{A}) and experiments (\textit{B}). Black symbols in (\textit{B}) and (\textit{C}) denote predictions of Eq.~6 with $N_+$ and $N_-$ as input from simulations. Shannon entropy $H$ of the time-forward sequence of states $P[\textbf{X}]$ as a function of $\theta$ from minimal simulations (\textit{C}) and experiments (\textit{D}). All shaded regions denote measurement standard deviations.
    }
    \label{fig:figure4_SI}
\end{figure*}

We measure $\sigma$ directly from filtered trajectories using recently introduced compression-based KLD estimator \cite{ro2022model} for our minimal simulations and experiments (Figs.~\ref{fig:figure4_SI}\textit{A} and \textit{B}). Specifically, we take a time-forward trajectory $\textbf{X}$ and divide it in two equal parts $\textbf{X}_\textbf{1}$ and $\textbf{X}_\textbf{2}$. We then use the difference of the cross-parsing complexity between $\textbf{X}_\textbf{1}$ and $\textbf{X}_\textbf{2}^\textbf{R}$ and between $\textbf{X}_\textbf{1}$ and $\textbf{X}_\textbf{2}$ to estimate $\sigma$. To confirm that our filtering procedure retained the essential time-irreversibility, we also measure $\sigma$ directly before and after filtering $\textbf{X}$ in minimal simulations. As shown in Fig.~\ref{fig:figure4_SI}\textit{A}, $\sigma$ remains the same with or without filtering. To obtain good estimates of $\sigma$ using compression-based KLD estimator, long time-series trajectories are typically required \cite{ro2022model}. For any given $\theta$, to get good experimental estimate of $\sigma$, we concatenate filtered time-series for different experimental runs into one single trajectory, which we then use to measure KLD. This method of getting long experimental time-series trajectories has also been used before \cite{tan2021scale}.

To calculate $\sigma$ analytically, we assume $\textbf{X}$ and $\textbf{X}^\textbf{R}$ follow Bernoulli distributions under the assumption that the crossing events of bacteria at low concentrations are uncorrelated. Nevertheless, since the dipolar flow induced by a moving bacterium is long-ranged and decays slowly with distance $r$ as $1/r^2$ \cite{drescher2011fluid}, it is necessary to independently check the assumption. From the set of all Boolean-valued probability distributions with a given probability parameter $p$, the Bernoulli distribution has the maximum Shannon entropy, since it is maximally random. If an unknown test distribution has Shannon entropy $H_{test}$, then $H_{test} \leq H_{Ber}$, where $H_{Ber}$ is the Shannon entropy of a Bernoulli distribution having the same $p$ value as the test distribution. We use a compression-based method \cite{ro2022model} to estimate Shannon entropy of $P[\textbf{X}]$, $H_{P[\textbf{X}]}$, directly from the time-series trajectories of our experiments and minimal simulations. As shown in Figs.~\ref{fig:figure4_SI}\textit{C} and \textit{D}, $H_{P[\textbf{X}]}$ from both experiments and simulations show good agreement with $H_{Ber}$. The result justifies the assumption of bacterial motion being uncorrelated, and consequently $P[\textbf{X}]$ being Bernoulli distributed.

\subsubsection{Relation between time-irreversibility and fluxes}

The KLD, $\sigma$, between two Bernoulli distributions with parameters $p$ and $1-p$ is given by:
\begin{equation}
    \sigma = p \ln \left( \frac{p}{1-p} \right) + (1-p) \ln \left( \frac{1-p}{p} \right).
    \label{eq:kld_ber}
\end{equation}
Since the probability density functions for time-forward and reverse dynamics $P[\textbf{X}]$ and $P[\textbf{X}^\textbf{R}]$ take the value $+v$ with a probability $p = {N}_{+}/({N}_{+}+{N}_{-})$ and $1-p = {N}_{-}/({N}_{+}+{N}_{-})$ respectively, Eq.~\ref{eq:kld_ber} leads to Eq.~6 of the main text.

To relate $\sigma$ with the normalized flux per particle $\eta_r$, we note that $\eta_r=({N}_{+}-{N}_{-})/({N}_{+}+{N}_{-})=2p-1$. Substituting $p=(\eta_r+1)/2$ in Eq.~\ref{eq:kld_ber}, we obtain
\begin{equation}
    \sigma = \eta_r \ln \frac{1+\eta_r}{1-\eta_r} = 2 \eta_r^2 + \frac{2}{3} \eta_r^4 + \mathcal{O}(\eta_r^6),
    \label{eq:kld_flux}
\end{equation}
where the second equality follows from the Taylor expansion of $\sigma$ around $\eta_r=0$. As evident from Eq.~\ref{eq:kld_flux}, when $\eta_r \to 0$, $\sigma \sim {\eta_r^2}$ (Fig.~3\textit{C} inset).

\subsubsection{Extractable work} 

Consider a particle moving at a constant velocity $v_{0}$ in a viscous fluid due to an internal driving force $F$. A small external force $F_{ext}$ is then applied to the particle opposite to the direction of $v_0$. The power extracted against $F_{ext}$ is simply 
\begin{equation}
    P(F_{ext}) = F_{ext}v_{ext},
    \label{eq:P_ext}
\end{equation} 
where $v_{ext}$ is the new steady-state velocity of the particle. For a small enough $F_{ext}$, $v_{ext}$ can be expanded to linear order in $F_{ext}/F_{stall}$, where the stall force $F_{stall}$ is the value of $F_{ext}$ that balances the internal driving force of the particle to give $v_{ext}=0$, i.e., $F_{stall} = F$. Thus, $v_{ext} = v_0 (1-F_{ext}/F) + \mathcal{O} ((F_{ext}/F)^2)$. Plugging $v_{ext}$ into Eq.~\ref{eq:P_ext} and maximizing $P(F_{ext})$ with respect to $F_{ext}$ gives the maximal extractable power $P = F v_{0}/4$. Finally, in the overdamped limit when the inertia is negligible, $v_0 = F/\mu$, we have
\begin{equation}
    P = \frac{F^2}{4\mu},
    \label{eq:p_f}
\end{equation}
where $\mu$ is the motility of the particle. 

We apply the above generic consideration in our study. Rectified bacterial motion provides a non-zero time-averaged driving force, $F=Ky_c$ to the colloid against the pulling of the harmonic trap, where $K$ is the harmonic trap constant and $y_c$ is the time-averaged $y$-position of the colloid, which is non-zero due to a net mass (and momentum) flux in the $+y$ direction. The harmonic trap can be thought of as an external load stalling the colloid. The no-load velocity then, is the velocity with which the colloid would move if $F$ alone was applied to it in the absence of an harmonic trap. Thus, Eq.~\ref{eq:p_f} applies directly to our study with $F = Ky_c$.

\subsubsection{Relation between time-irreversibility, fluxes, and extractable work} 

We derive the relation beween fluxes, time irreversibility, and extractable work in the ideal scenario, where we trap a colloid, coupled weakly to bacterial motion, in a harmonic potential at the common tip of two oppositely-facing funnels with the same angle $\theta=90\degree$ and different wall lengths (Fig.~4\textit{A}). To relate extractable power to fluxes, we need to relate the time-averaged driving force $F=Ky_c$ to the normalized flux per particle $\eta_r$. As $y_c$ is the time-averaged $y$-position of the colloid, it can be calculated as:
\begin{equation}
    y_c = \frac{y_{u}t_{u}\mathcal{N}_{u} + y_{d}t_{d}\mathcal{N}_{d} + y_{rest}t_{rest}}{t_{tot}}, 
    \label{eq:y_c}
\end{equation}
where $y_{u}$ ($y_{d}$) is the time-averaged $y$-position of the colloid per bacterium-colloid collision in the $+y$ $(-y)$ direction, $t_u$ $(t_d)$ is the average time for which the colloid is displaced from its rest position $y_{rest}$ in the $+y$ $(-y)$ direction per bacterium-colloid collision, $\mathcal{N}_u$ $(\mathcal{N}_d)$ is the total number of bacterium-colloid collisions in the $+y$ $(-y)$ direction over a time interval $t_{tot}$, $t_{rest}$ is the time interval over which the colloid stays at the rest position. As $y_{rest} = 0$ at the center of the harmonic trap, $y_{rest}t_{rest} = 0$. Furthermore, since the two funnels have the same angle $\theta = 90\degree$ in the ideal scenario, $t_u=t_d$, and $y_{u}=-y_{d}$. Hence, Eq.~\ref{eq:y_c} can be simplified as,
\begin{equation}
    y_c = \frac{y_{u}t_{u}}{t_{tot}} (\mathcal{N}_{u} - \mathcal{N}_{d}), 
    \label{eq:y_c_simplified}
\end{equation}
Using Eq.~\ref{eq:y_c_simplified}, $F$ can be written as,
\begin{equation}
    F = \frac{Ky_{u}t_{u}}{t_{tot}} (\mathcal{N}_{u} - \mathcal{N}_{d}), 
    \label{eq:F_j}
\end{equation}

Note that $\eta_r \equiv ({N}_{+}-{N}_{-})/({N}_{+}+{N}_{-}) = (\mathcal{N}_{u}-\mathcal{N}_{d})/(\mathcal{N}_{u}+\mathcal{N}_{d})$. Defining $\mathcal{N}=\mathcal{N}_{u}+\mathcal{N}_{d}$ as the total number of bacteria crossing through the funnel tip and inserting Eq.~\ref{eq:F_j} into Eq.~\ref{eq:p_f}, we have,
\begin{equation}
    P = \left[ \frac{1}{4\mu} \left( \frac{Ky_{u}t_{u}}{t_{tot}}\right)^2 \right] \mathcal{N}^2 \eta_r^2 \equiv c\mathcal{N}^2 \eta_r^2, 
    \label{eq:P_flux}
\end{equation}
with all the system-specific parameters absorbed in the constant $c$. 

Finally, combining Eqs.~\ref{eq:kld_flux} and \ref{eq:P_flux} and approximating $\sigma$ using the first term in the Taylor series as $\eta_r \to 0$, we have
\begin{equation}
   P = c\mathcal{N}^2 \eta_r^2 \underset{\eta_r \to 0}{=} \frac{1}{2}c\mathcal{N}^2\sigma,
   \label{eq:P_flux_sigma}
\end{equation}
which is Eq.~7 of the main text.

\section{Description of Supplemental Movies}

\begin{itemize}

\item \textbf{Supplemental Movie 1:} Rectification of {\it E. coli} motions by a PDMS funnel wall of angle $\theta = 50\degree$ (Fig.~1\textit{A}). 

\item \textbf{Supplemental Movie 2:} Rectification of {\it E. coli} motions by a PDMS funnel wall of angle $\theta = 150\degree$ (Fig.~1\textit{B}).

\item \textbf{Supplemental Movie 3:} Rectification of {\it E. coli} motions by a colloidal funnel wall of angle $\theta = 50\degree$. The bacterium, highlighted in yellow, approaches the tip colloid from the $+y$ direction (Figs.~5\textit{D}, Top row). The wall is constructed by harmonically trapping 4-$\mu$m-diameter colloids with optical tweezers. The tip colloid is held weakly compared to wall colloids.

\item \textbf{Supplemental Movie 4:} Rectification of {\it E. coli} motions by a colloidal funnel wall of angle $\theta = 50\degree$. The bacterium, highlighted in yellow, approaches the tip colloid from the $-y$ direction (Figs.~5\textit{D}, Bottom row). The wall is constructed by harmonically trapping 4-$\mu$m-diameter colloids with optical tweezers. The tip colloid is held weakly compared to wall colloids.

\end{itemize}

\bibliographystyle{apsrev4-2}

\bibliography{ref.bib}

\pagebreak

\begin{figure*}[htpb]
    \centering
    \includegraphics[width=0.5\linewidth]{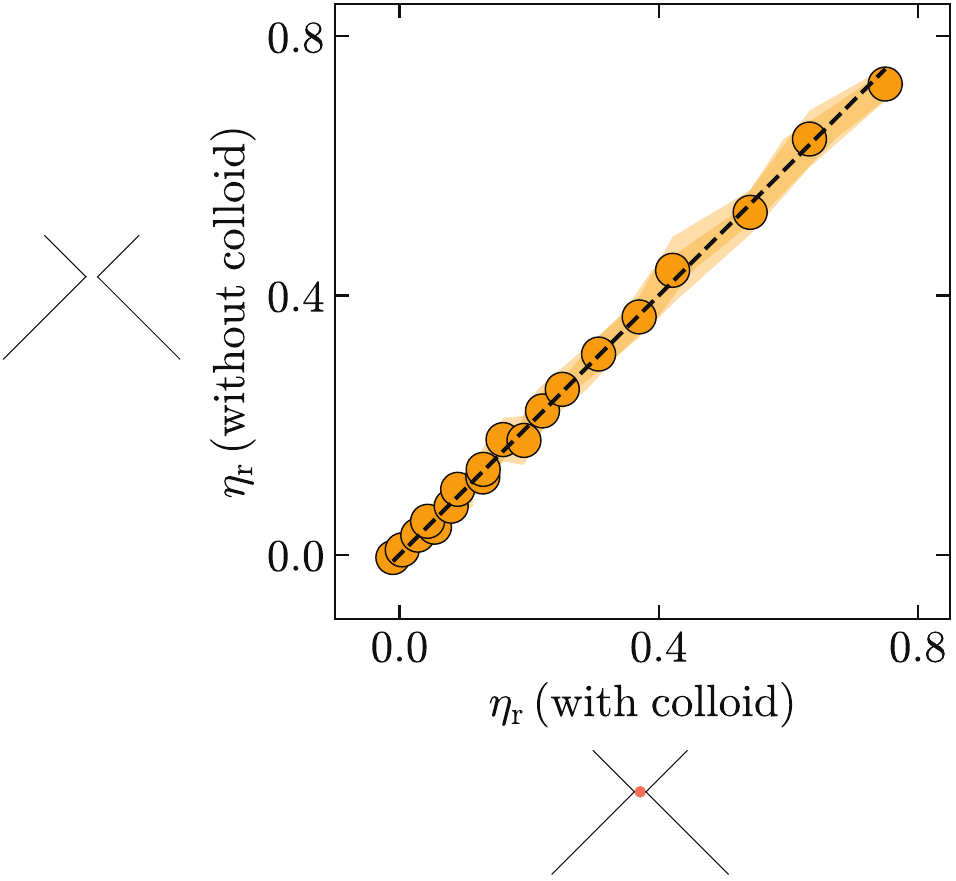}
    \caption{\small Weak-coupling in the ideal scenario. Normalized flux per particle $\eta_r$ for the system shown in Fig.~4\textit{A} with a harmonically trapped colloid at the common tip of two oppositely-facing funnel rectifiers versus $\eta_r$ for the same system without the trapped colloid. The black dashed line indicates $y=x$. The result shows that particle flux is not affected by the presence of the colloid. Shaded regions denote measurement standard deviations. 
    }
    \label{fig:figure5_SI}
\end{figure*}

\pagebreak

\begin{figure*}[!htpb]
    \centering
    \includegraphics[width=\linewidth]{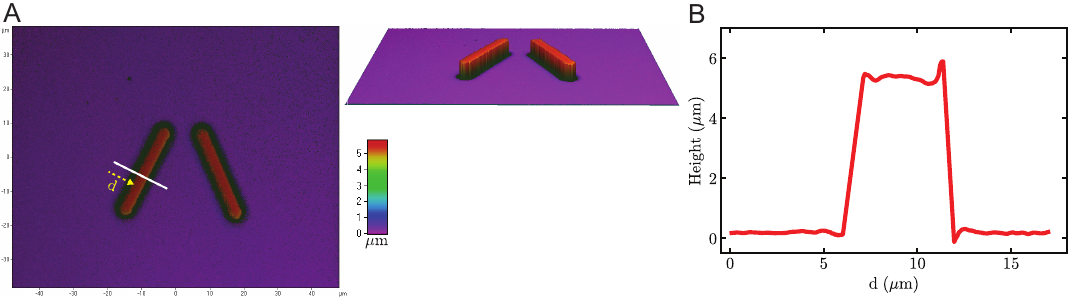}
    \caption{\small {PDMS channel characterization.} (\textit{A}) Optical profiling images of an exemplar PDMS channel with a funnel rectifier at the center. (\textit{B}) Height profile of the PDMS channel plotted as a function of distance $d$ along the white line across the funnel rectifier shown in (\textit{A}). 
    }
    \label{fig:figure6_SI}
\end{figure*}